\documentclass{egpubl}
\usepackage{sca2026}
 
\SpecialIssuePaper         

\CGFccby

\usepackage{tabularray}
\UseTblrLibrary{booktabs}
\usepackage{xcolor}

\usepackage[T1]{fontenc}
\usepackage{dfadobe}
\usepackage{cite}
\BibtexOrBiblatex
\electronicVersion
\PrintedOrElectronic

\ifpdf
  \usepackage[pdftex]{graphicx}
\else
  \usepackage[dvips]{graphicx}
\fi

\usepackage{egweblnk}
\usepackage{booktabs}
\usepackage{amsmath,amssymb}
\usepackage{xcolor}
\usepackage{siunitx}
\title[Closing Trajectories]{Closing Trajectories: Equation-Free Cyclic Animation\\ via Koopman Surrogates}

\makeatletter
\def\@fnsymbol#1{\ifcase#1\or *\or \dagger\or \ddagger\or
 \mathchar "278\or \mathchar "27B\or \|\or **\or \dagger\dagger\or
 \ddagger\ddagger\else ***\fi\relax}
\makeatother

\author[Huang]
{\parbox{\textwidth}{\centering
Shixun Huang$^{1,2}$\thanks{\raggedright Equal contribution.}\orcid{0009-0002-2463-1133}
\quad
Siyuan Chen$^{2}$\footnotemark[1]\thanks{\raggedright Corresponding authors:
\texttt{csy0105@student.ubc.ca},\\
\texttt{peter.chen@ubc.ca}.}\orcid{0009-0009-4309-862X}
\quad
Yue Chang$^{1}$\orcid{0000-0002-2587-827X}
\quad
Zhecheng Wang$^{1}$\orcid{0000-0003-4989-6971}
\quad
Peter Yichen Chen$^{2}$\footnotemark[2]\orcid{0000-0003-1919-5437}
\\[2pt]
$^{1}$University of Toronto, Toronto, Canada
\quad
$^{2}$University of British Columbia, Vancouver, Canada
}}

\newcommand{\Revision}[1]{#1}

\begin{document}

\teaser{
  \centering
    \includegraphics[width=\linewidth]{figures_type3_fixed/teaser.pdf}
  \caption{Given an input non-cyclic sequence (left top), we edit the trajectory into a cyclic one (left bottom) while retaining the characteristic dynamics observed in the input. Our method applies consistently across different physical systems (right), including cloth, deformable bodies, and fluids. The edited results exhibit seamless temporal looping while maintaining the characteristic dynamics of the input.}
  \label{fig:teaser}
}

\maketitle

\begin{abstract}
Cyclic animation is widely used in computer graphics and interactive content. It supports seamless playback in games, VR, and interactive simulation, where short clips must repeat smoothly over long durations. \Revision{Converting an observed non-cyclic trajectory into a seamless loop} is challenging because the endpoint states of the observed sequence rarely match exactly, and the governing equations of the underlying system are often unavailable. We therefore propose an equation-free framework that identifies a Koopman surrogate from the observed trajectory and computes a cyclic trajectory by applying a Fourier-parameterized, time-varying \Revision{control input} under a hard temporal periodicity constraint. The resulting formulation reduces cyclic synthesis to a linearly constrained quadratic program that can be solved efficiently through a structured KKT system. Our method is applicable to a diverse range of examples, including N-body systems, cloth, deformable objects, shallow water, etc.
\end{abstract}

\section{Introduction}

A cyclic animation is an animation sequence that repeats seamlessly over time. It is valuable for interactive applications that must run for long durations, because a short cyclic clip can be repeated indefinitely without manual reauthoring. Cyclic sequences also serve as reusable building blocks in animation systems, enabling consistent behaviors across scenes and facilitating compositional editing. As a result, the ability to synthesize high-quality cyclic animation from general input sequences is practically important in computer graphics and simulation workflows.

\begin{figure*}[t]
    \centering
    \includegraphics[width=0.98\textwidth]{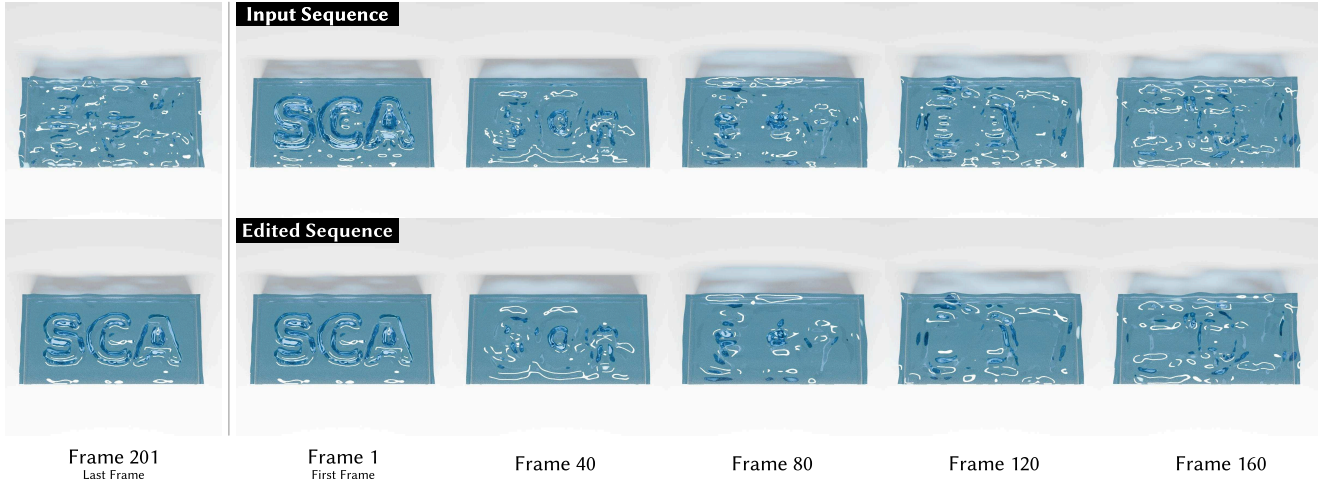}
    \caption{SCA-shaped Pattern. Top: the input sequence is non-cyclic and does not recover the initial SCA-shaped free-surface pattern at the last frame. Bottom: our method restores the initial SCA-shaped pattern while preserving the overall wave evolution, producing a seamless loop.}
    \label{fig:shallow_water_sca}
\end{figure*}

Formally, a cyclic animation is a trajectory in state space that returns to its initial state after one period. However, converting a general input sequence into such an animation is challenging. A visually closed loop may still contain abrupt or unnatural changes over time: simply forcing the last frame to match the first may introduce distortions that disrupt \Revision{the characteristic dynamics observed in the input}. In other words, cyclicity is not only a geometric endpoint condition, but also an editing objective that should preserve \Revision{the dominant dynamics of the observed trajectory}.

Existing methods for synthesizing cyclic animation generally fall into two categories. Interpolation-based methods such as blending \cite{Kovar2002motion,Park2002online} and frame stitching \cite{Heck2007motion,Mizuguchi2001data} enforce periodic closure directly on the observed sequence, but often do so at the cost of physical plausibility. On the other hand, physics-based methods \cite{Witkin1988spacetime,Barbic2009control,Jia2023physical} can produce physically plausible results with given physics constraints, but they typically rely heavily on access to governing equations, energy models, or simulators, and often involve expensive nonlinear optimization. This dependence limits their applicability in settings where only a trajectory is given, such as precomputed simulation outputs, legacy animation data, or captured motion sequences for which the original simulator or physical model is unavailable.



In this work, we propose an \emph{equation-free} framework for synthesizing cyclic animation from an observed trajectory. We first identify a Koopman operator from the observed trajectory and use it as a surrogate for the unknown dynamics. We then solve for the minimal \Revision{control input} that steers the trajectory toward a cyclic one. This formulation seeks a cyclic trajectory that remains close to the observed sequence while using only the minimal \Revision{control input} needed to enforce periodicity, which helps preserve a \Revision{visually} plausible temporal evolution. \Revision{The proposed method is primarily an \textbf{animation-editing} framework rather than a physically faithful simulation approach.}

Our approach is fully data-driven and does not require an explicit physical model. The resulting problem becomes a linearly constrained quadratic program with a structured KKT system, making the method computationally efficient. By editing the trajectory through a learned surrogate dynamics rather than direct interpolation, the method aims to preserve the temporal structure of the observed sequence while enforcing exact cyclic closure. 


In summary, this work makes the following contributions:
\begin{itemize}
    \item We introduce a purely data-driven, equation-free framework for synthesizing cyclic animation. Given an observed physical trajectory, our method fits a Koopman surrogate to the input trajectory and computes a minimal control that steers it toward exact cyclic closure, without knowledge of the underlying dynamics equation. 
    \item We show that the resulting formulation reduces to a linearly constrained quadratic program that can be solved efficiently through a structured Karush--Kuhn--Tucker (KKT) system.
    
\end{itemize}

\section{Related Work}


\textbf{Koopman operator methods.}
Koopman-based approaches model nonlinear dynamics through linear operators in lifted spaces and have been extensively studied for the analysis and control of dynamical systems \cite{koopman1931hamiltonian, Proctor2016dynamic,Proctor2018generalizing}. In computer graphics, \cite{deAguiar2010Stable} proposed a method to linearize the operator for cloth simulation conditioned on an animation skeleton, whereas our approach does not require a skeleton as input.

Our method adopts Dynamic Mode Decomposition (DMD) \cite{schmid2010dynamic} to approximate the Koopman operator. Interest in DMD has grown within the graphics community in recent years. It has been shown that DMD can be used for fast fluid simulation \cite{chen2025dmd} and deformable simulation \cite{chang2026lowrank}. In both cases, the authors leveraged the linearity of the operator to enable editing and inverse problems. However, none of the aforementioned methods discuss how to generate cyclic animations using the approximated Koopman operator. We show, for the first time, that the DMD framework can be applied to generate cyclic animations. Relatedly, Dynamic Mode Decomposition with control extends this framework to systems with prescribed input signals \cite{Proctor2016dynamic}. \Revision{Unlike DMD with control, which requires observed state-control pairs to identify a controlled linear system, our method observes only a trajectory. We first learn a reduced DMD-based surrogate from this trajectory alone, then solve a constrained editing problem that computes a minimal corrective control to close the motion into a cyclic orbit. Control in our formulation is thus not a measured physical input but an editing variable introduced to enforce cyclic closure while preserving the observed behavior.}


\textbf{Reduced-order modeling.}
Reduced-order models are widely used to represent high-dimensional simulations in a compact space~\cite{barbic2005realtime,Kim2009skipping,de2010stable}, and have recently been extended through neural approaches~\cite{chen2023crom,Chang2023liCROM, Modi:2024:Simplicits,Chang2025shape,chang2026lowrank}. 
By projecting the full space configuration onto a small number of dominant modes, these methods \cite{Barbic2009control,Barbic2012interactive,Chen2024fluid} make simulation and editing significantly more efficient while preserving the main behaviors of the system. Prior work \cite{Harmon2013subspace,Neumann2013sparse,Lan2024efficient,Benchekroun2025force} has further extended this idea with local enrichment and nonlinear reduced representations to better capture complex deformation patterns. Our work is related in spirit because we also operate in a low-dimensional subspace learned from the observed trajectory. The key difference is that traditional reduced-order methods typically rely on an explicit physical model or simulator, whereas our setting is equation-free. We do not reduce a known simulation model; instead, we learn a reduced surrogate directly from data and use it to impose cyclicity while staying close to \Revision{the characteristic dynamics observed in the input trajectory}. Our interactive extension further shows that our method can be combined with localized control bases for user-guided local editing.

\textbf{Data-only methods.} Prior work has explored motion generation from a single image~\cite{tedla2025generating} \Revision{\cite{Li_2024_CVPR}} or video~\cite{kansy2025reenact}, with some methods enabling personalization through generative priors~\cite{abdal2025dynamic}. Several approaches also operate directly on 3D motion data, such as motion manifolds~\cite{starke2022deepphase} and motion generation from limited data~\cite{li2022ganimator}. These methods, however, do not impose structure motivated by underlying dynamics. In contrast, we introduce a structured temporal prior by approximating the system evolution with a linear operator. This assumption reflects the expectation that the data arises from a coherent dynamical process, and may be complementary to existing approaches. 

\section{Background}

\subsection{Cyclic Animation}\label{sec:background_koopman_theory}

Consider a continuous-time dynamical system with state \(x(t)\in\mathcal{X}\), governed by
\begin{equation}\label{eq:continuous_dynamics}
\frac{d x(t)}{dt}=f(x(t)),
\end{equation}
where \(f\) denotes the underlying dynamics. Let \(F^t:\mathcal{X}\to\mathcal{X}\) denote the corresponding flow map, so that \(F^t(x(0))\) gives the state reached after evolving the system for time \(t\) from initial state \(x(0)\). A trajectory is cyclic with period \(T\) if
\begin{equation}\label{eq:cyclic_definition_continuous}
x(T)=x(0),
\end{equation}
or equivalently,
\begin{equation}\label{eq:cyclic_definition_flow}
F^{T}(x(0))=x(0).
\end{equation}

When the governing dynamics are known, periodic closure can be formulated directly with respect to the underlying system. In practice, however, observed trajectories are often more readily available than the underlying governing dynamics. We therefore learn a surrogate evolution model directly from the observations and enforce periodicity with respect to that surrogate.

\subsection{Koopman Surrogate}
\label{sec3:koopman_surrogate}
\begin{figure*}[ht!]
    \centering
    \includegraphics[width=\textwidth]{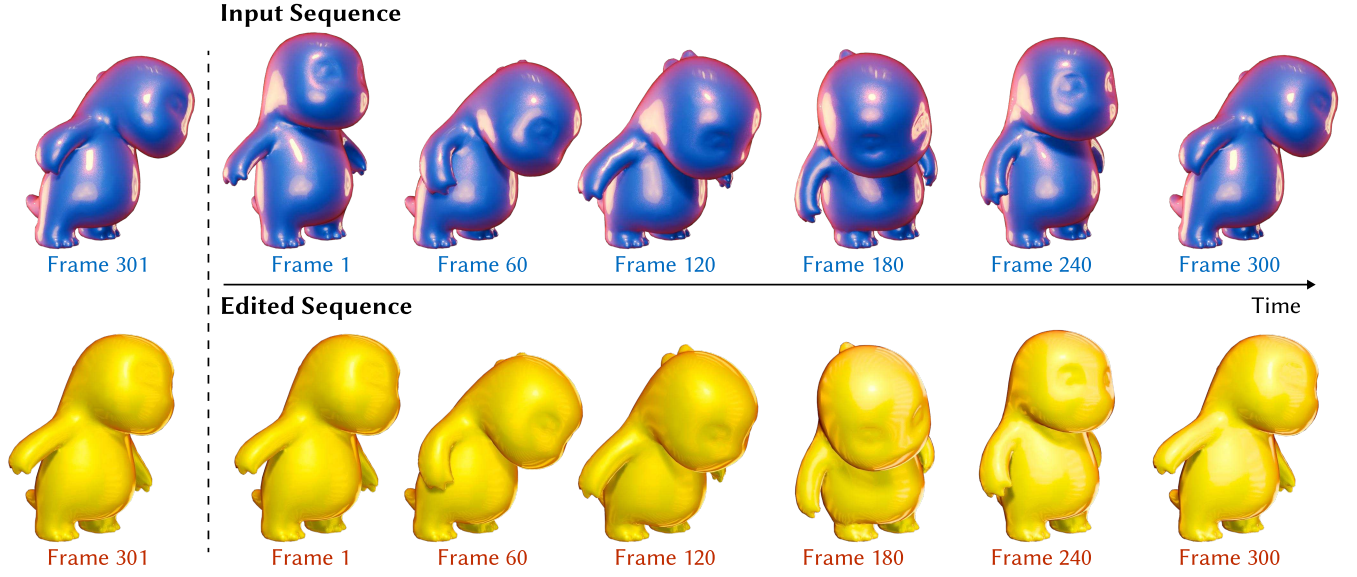}
    \caption{Cute dragon. Top: the input deformable sequence is non-cyclic, resulting in an open trajectory whose final frame does not return to the initial configuration. Bottom: our method transforms the sequence into a cyclic one over the same duration, closing the trajectory while preserving the overall deformation pattern. The resulting sequence transitions smoothly from the final frame back to the initial frame, forming a seamless loop.}
    \label{fig:cute_dragon}
\end{figure*}

To model the observed dynamics without access to the governing equations, we use a data-driven surrogate inspired by Koopman theory \cite{koopman1931hamiltonian}. The Koopman viewpoint describes nonlinear evolution through a linear operator acting on observables of
the state, \Revision{ where an observable is a function $(g:\mathcal{X}\to\mathbb{R}^p)$ that maps a state to quantities of interest.} For a flow map \(F^t\) and an observable \(g\), the Koopman operator \(\mathcal{K}^t\) is defined by
\begin{equation}
(\mathcal{K}^{t}g)(x)=g\bigl(F^{t}(x)\bigr).
\end{equation}

In practice, we are given a finite trajectory of sampled animation states, where each frame is represented by a spatially discretized state vector \(\mathbf{x}_t \in \mathbb{R}^n\), \Revision{i.e., the concatenated degrees of freedom of the discretized object such as vertex positions}, at discrete time step $t$, for \(t=1,\dots,T\).  From these sampled states, we construct a finite-dimensional approximation \cite{schmid2010dynamic} directly from the trajectory data.

Given the sampled states \(\{\mathbf{x}_t\}_{t=1}^{T}\) where each $\mathbf{x}_t \in \mathbb{R}^n$. We define the snapshot matrices
\begin{equation}
\mathbf{X}=
\begin{bmatrix}
\mathbf{x}_1 & \mathbf{x}_2 & \cdots & \mathbf{x}_{T-1}
\end{bmatrix},
\qquad
\mathbf{X}'=
\begin{bmatrix}
\mathbf{x}_2 & \mathbf{x}_3 & \cdots & \mathbf{x}_{T}
\end{bmatrix}.
\end{equation}
A standard finite-dimensional approximation of the discrete Koopman operator is obtained by solving the least-squares problem
\begin{equation}
\mathbf{K}
=
\arg\min_{\mathbf{A}}
\|\mathbf{X}'-\mathbf{A}\mathbf{X}\|_F^2,
\label{eq:koopman_least_squares}
\end{equation}
where \(\mathbf{K}\in\mathbb{R}^{n\times n}\).
This yields the one-step surrogate model
\begin{equation}
\mathbf{x}_{t+1}\approx\mathbf{K}\mathbf{x}_t.
\label{eq:one_step_evolution}
\end{equation}

Although the underlying dynamics and the associated flow map introduced earlier are unknown and may be nonlinear, prior work has shown that Koopman surrogates can provide useful approximations for complex physical systems in \cite{chen2025dmd,chang2026lowrank}. In our setting, we do not require an exact recovery of the true dynamics, but rather a data-driven surrogate that captures the sampled evolution sufficiently well to support cyclic trajectory synthesis. This surrogate is particularly useful because its linear structure enables a closed-form rollout and ultimately leads to a linearly constrained quadratic program that can be solved via a KKT system.


\subsection{Low-Rank Koopman Approximation in a Reduced Space}
\label{sec:low_rank_koopman}

For sampled animation trajectories, however, the spatially discretized state vectors can be very high-dimensional, making a dense operator in the full state space expensive to learn and evolve, and potentially sensitive to noise. We therefore construct the surrogate in a reduced space spanned by the dominant modes of the observed trajectory~\cite{schmid2010dynamic}.

Given the rank-$r$ truncated singular value decomposition of 
\(\mathbf{X}\in\mathbb{R}^{n\times (T-1)}\),
\begin{equation}
\mathbf{X}\approx \mathbf{U}_r\boldsymbol{\Sigma}_r\mathbf{V}_r^\top,
\end{equation}
where 
\(\mathbf{U}_r\in\mathbb{R}^{n\times r}\),
\(\boldsymbol{\Sigma}_r\in\mathbb{R}^{r\times r}\), and
\(\mathbf{V}_r\in\mathbb{R}^{(T-1)\times r}\),
the columns of \(\mathbf{U}_r\) form a low-dimensional basis capturing the main variation of the observed trajectory. We then define reduced coordinates
\begin{equation}
\mathbf{z}_t=\mathbf{U}_r^\top\mathbf{x}_t \in \mathbb{R}^r.\label{eq:latent_coordinates}
\end{equation}
Let
\begin{equation}
\mathbf{Z}=
\begin{bmatrix}
\mathbf{z}_1 & \mathbf{z}_2 & \cdots & \mathbf{z}_{T-1}
\end{bmatrix},
\qquad
\mathbf{Z}'=
\begin{bmatrix}
\mathbf{z}_2 & \mathbf{z}_3 & \cdots & \mathbf{z}_{T}
\end{bmatrix}.
\end{equation}
We then fit the reduced surrogate \(\hat{\mathbf{K}}\in\mathbb{R}^{r\times r}\) by solving
\begin{equation}
\hat{\mathbf{K}}
=
\arg\min_{\mathbf{A}}
\|\mathbf{Z}'-\mathbf{A}\mathbf{Z}\|_F^2,
\label{eq:reduced_koopman}
\end{equation}
so that the reduced-space evolution is approximated by
\begin{equation}
\mathbf{z}_{t+1}\approx\hat{\mathbf{K}}\mathbf{z}_t.
\label{eq:latent_dynamics}
\end{equation}
\Revision{
Equivalently, this reduced evolution induces a projected full-state operator,
\begin{equation}
\mathbf{x}_{t+1}
\approx
\mathbf{U}_r \hat{\mathbf{K}} \mathbf{U}_r^\top \mathbf{x}_t,
\end{equation}
which shows that the reduced Koopman surrogate corresponds to projecting the full-state evolution onto the subspace spanned by $\mathbf{U}_r$.} The reduced model provides a compact surrogate for the observed trajectory. In the remainder of the paper, we formulate the cyclic synthesis problem directly in the reduced space.

\section{Formulation}
\subsection{Full-Space Cyclic Control}

As discussed in Sec.~\ref{sec:background_koopman_theory}, cyclicity in the continuous setting is characterized by closure under the underlying flow map. In practice, however, we work with the learned discrete Koopman surrogate rather than the unknown governing dynamics. If we simply roll out the learned surrogate from the observed trajectory, \Revision{it only approximates the projected sampled evolution and does not, in general, satisfy a cyclic closure condition.} To synthesize a cyclic trajectory, we therefore introduce a time-dependent control term that steers the surrogate dynamics away from the original rollout while keeping the correction as small as possible, in the spirit of prior keyframe-guided control formulations \cite{Barbic2009control}.

Let \(\mathbf{x}_t \in \mathbb{R}^n\) denote the observed sampled state at time step \(t\), and let \(\tilde{\mathbf{x}}_t \in \mathbb{R}^n\) denote the corrected trajectory to be optimized. Using the learned full-space surrogate \(\mathbf{K}\in\mathbb{R}^{n\times n}\), we model the corrected evolution as
\begin{equation}
\tilde{\mathbf{x}}_{t+1}
=
\mathbf{K}\tilde{\mathbf{x}}_t+\mathbf{c}_t,
\qquad t=1,\ldots,T,
\end{equation}
where \(\mathbf{c}_t\in\mathbb{R}^n\) is a full-space control input. Periodicity is enforced by the discrete closure condition
\begin{equation}
\tilde{\mathbf{x}}_{T+1}=\tilde{\mathbf{x}}_1.
\end{equation}
This leads to the full-space cyclic control problem
\begin{equation}
\begin{aligned}
\min_{\tilde{\mathbf{x}}_1,\{\mathbf{c}_t\}}
\;&
w_{\mathrm{full}}\sum_{t=1}^{T}\|\tilde{\mathbf{x}}_t-\mathbf{x}_t\|_2^2
+
w_c\sum_{t=1}^{T}\|\mathbf{c}_t\|_2^2 \\
\text{s.t.}\;&
\tilde{\mathbf{x}}_{t+1}
=
\mathbf{K}\tilde{\mathbf{x}}_t+\mathbf{c}_t,
\qquad t=1,\ldots,T, \\
&
\tilde{\mathbf{x}}_{T+1}=\tilde{\mathbf{x}}_1.
\end{aligned}
\label{eq:optimization_problem_control_cyclic_full}
\end{equation}
where \(w_{\mathrm{full}}, w_c > 0\) balance fidelity to the observed trajectory and the magnitude of the control input.

\subsection{Reduced-Space Formulation}

For the high-dimensional trajectories considered here, solving \eqref{eq:optimization_problem_control_cyclic_full} directly in the full state space can be computationally expensive. We therefore consider a reduced-space version of the same formulation, based on the reduced surrogate introduced in Sec.~\ref{sec:low_rank_koopman}. This preserves the same control-based structure while significantly reducing the number of unknowns.

Let \(\mathbf{z}_t\in\mathbb{R}^{r}\) denote the observed reduced state at time step \(t\), and let \(\tilde{\mathbf{z}}_t\in\mathbb{R}^{r}\) denote the reduced trajectory to be optimized. Using the learned reduced surrogate \(\hat{\mathbf{K}}\in\mathbb{R}^{r\times r}\), we model the corrected evolution as
\begin{equation}
\tilde{\mathbf{z}}_{t+1}
=
\hat{\mathbf{K}}\tilde{\mathbf{z}}_t+\mathbf{u}_t,
\qquad t=1,\ldots,T,
\label{eq:discrete_controlled_dynamics}
\end{equation}
where \(\mathbf{u}_t\in\mathbb{R}^{r}\) is a reduced-space control input. The cyclic closure condition becomes
\begin{equation}
\tilde{\mathbf{z}}_{T+1}=\tilde{\mathbf{z}}_1.
\label{eq:discrete_periodic_closure}
\end{equation}

The reduced optimization problem becomes
\begin{equation}
\begin{aligned}
\min_{\tilde{\mathbf{z}}_1,\{\mathbf{u}_t\}}
\;&
w_{\mathrm{red}}\sum_{t=1}^{T}\|\tilde{\mathbf{z}}_t-\mathbf{z}_t\|_2^2
+
w_u\sum_{t=1}^{T}\|\mathbf{u}_t\|_2^2 \\
\text{s.t.}\;&
\tilde{\mathbf{z}}_{t+1}
=
\hat{\mathbf{K}}\tilde{\mathbf{z}}_t+\mathbf{u}_t,
\qquad t=1,\ldots,T, \\
&
\tilde{\mathbf{z}}_{T+1}=\tilde{\mathbf{z}}_1.
\end{aligned}
\label{eq:optimization_problem_control_cyclic}
\end{equation}
where \(w_{\mathrm{red}},w_u>0\) balance fidelity to the observed reduced trajectory and regularization of the reduced-space control input.
\begin{figure*}[]
    \centering
    \includegraphics[width=\textwidth]{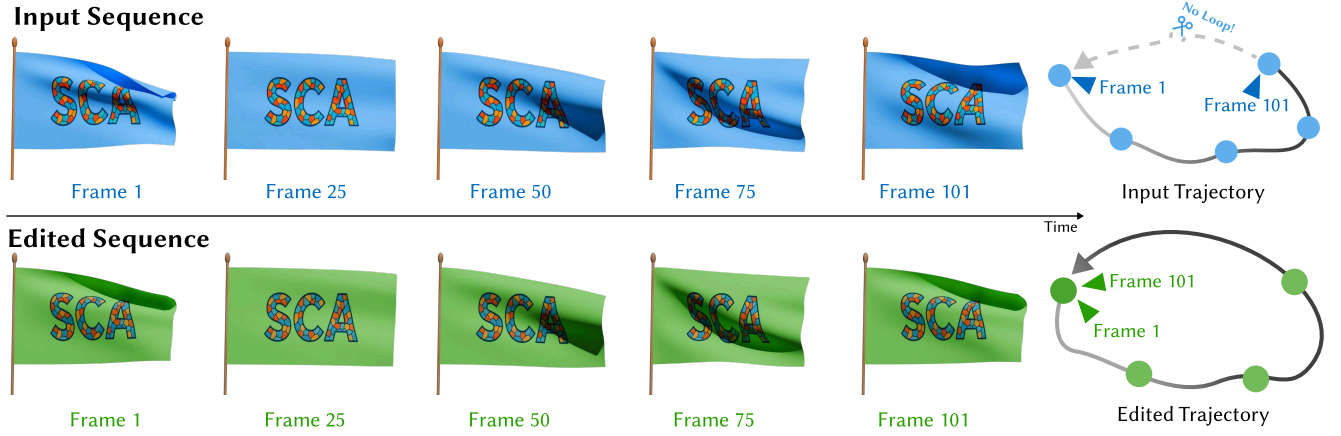}
    \caption{Flag-in-the-wind. Top: the input sequence is non-cyclic. Bottom: our method synthesizes a cyclic sequence of the same duration while preserving the dominant fluttering behavior. The trajectory on the right is closed smoothly, producing a seamless loop.}
    \label{fig:flag}
\end{figure*}

\section{Method}

Eq.~\eqref{eq:optimization_problem_control_cyclic} gives a natural reduced-space formulation for cyclic trajectory synthesis, but it is still written in recursive form through the surrogate dynamics. To obtain a solver-friendly formulation, we further exploit the linear structure of the learned Koopman surrogate and write the reduced trajectory explicitly in terms of the unknown initial state and control variables. This leads to a compact linearly constrained quadratic program, which can then be solved efficiently through its KKT system.

\subsection{Closed-Form Rollout in Reduced Space}

Starting from Eq.~\eqref{eq:discrete_controlled_dynamics} the reduced trajectory admits the closed-form expression
\begin{equation}
\tilde{\mathbf{z}}_{t}
=
\hat{\mathbf{K}}^{t-1}\tilde{\mathbf{z}}_1
+
\sum_{j=1}^{t-1}\hat{\mathbf{K}}^{t-1-j}\mathbf{u}_j,
\qquad t=1,\ldots,T.
\end{equation}
Imposing periodicity yields
\begin{equation}
\tilde{\mathbf{z}}_1
=
\hat{\mathbf{K}}^{T}\tilde{\mathbf{z}}_1
+
\sum_{j=1}^{T}\hat{\mathbf{K}}^{T-j}\mathbf{u}_j.
\end{equation}

\subsection{Fourier Parameterization of the \Revision{Control Input}}

To obtain a compact temporal representation of the \Revision{control input}, we consider parameterizing it using a truncated Fourier basis. We therefore write
\begin{equation}
\mathbf{u}_t = \Gamma\mathbf{s}_t,
\qquad t=1,\dots,T,
\end{equation}
where \(\Gamma \in \mathbb{R}^{{r\times m}}\) is a coefficient matrix and
\begin{equation}
\mathbf{s}_t=
\begin{bmatrix}
\sin\frac{2\pi t}{T},\ \cos\frac{2\pi t}{T},\ \ldots,\
\sin\frac{2\pi \Revision{M}t}{T},\ \cos\frac{2\pi \Revision{M}t}{T}
\end{bmatrix}^{\top}\in\mathbb{R}^{m}
\end{equation}
collects the evaluations of the selected Fourier modes at time step \(t\), \Revision{where \(M\) is the number of retained harmonics and \(m=2\Revision{M}\) (or \(2\Revision{M}+1\) if a constant mode is included).}

This parameterization restricts the \Revision{control input} to be globally smooth and
periodic in time. It
also reduces the number of unknowns from \(rT\) to \(rm\) in the collapsed form
through \(\Gamma\). Although Fourier is not the only possible temporal basis, it
is a natural choice for cyclic trajectories defined on a periodic time grid. 

With this parameterization, the rollout becomes
\begin{equation}
\tilde{\mathbf{z}}_{t}
=
\hat{\mathbf{K}}^{t-1}\tilde{\mathbf{z}}_1
+
\sum_{j=1}^{t-1}\hat{\mathbf{K}}^{t-1-j}\Gamma \mathbf{s}_j,
\qquad t=1,\dots,T,
\end{equation}
and the cyclic boundary condition becomes
\begin{equation}
\tilde{\mathbf{z}}_1
=
\hat{\mathbf{K}}^{T}\tilde{\mathbf{z}}_1
+
\sum_{j=1}^{T}\hat{\mathbf{K}}^{T-j}\Gamma \mathbf{s}_j.
\end{equation}

\subsection{Linearly Constrained Quadratic Program}

We optimize \(\tilde{\mathbf{z}}_1\) and \(\Gamma\) by stacking unknowns as
\begin{equation}
\mathbf{q}
=
\begin{bmatrix}
\tilde{\mathbf{z}}_1\\
\mathrm{vec}(\Gamma)
\end{bmatrix}
\in\mathbb{R}^{r+rm},
\qquad
\mathbf{y}
=
\begin{bmatrix}
\mathbf{z}_1\\
\vdots\\
\mathbf{z}_T
\end{bmatrix}
\in\mathbb{R}^{rT}.
\end{equation}

Define
\begin{equation}
E_z:=\begin{bmatrix}I_r & 0_{r\times rm}\end{bmatrix},
\qquad
E_{\Gamma}:=\begin{bmatrix}0_{rm\times r} & I_{rm}\end{bmatrix},
\end{equation}
and recursively construct rollout matrices
\begin{equation}
R_1=E_z,
\qquad
R_{t+1}=\hat{\mathbf{K}}R_t+\left(\mathbf{s}_t^{\top}\!\otimes I_r\right)E_{\Gamma}.
\end{equation}
Then each reduced state is linear in \(\mathbf{q}\):
\begin{equation}
\tilde{\mathbf{z}}_t=R_t\mathbf{q}.
\end{equation}
Stacking all frames gives
\begin{equation}
\tilde{\mathbf{Z}}
=
\begin{bmatrix}
R_1\\
\vdots\\
R_T
\end{bmatrix}
\mathbf{q}
=
A\mathbf{q},
\qquad
A:=\begin{bmatrix}
R_1\\
\vdots\\
R_T
\end{bmatrix}.
\end{equation}

The hard closure constraint is
\begin{equation}
\Revision{A_{\mathrm{cl}}}\mathbf{q}=0,
\qquad
\Revision{A_{\mathrm{cl}}}:=R_{T+1}-R_1.
\end{equation}

For the \Revision{control input} regularization,
\begin{equation}
\sum_{t=1}^{T}\|\mathbf{u}_t\|_2^2
=
\sum_{t=1}^{T}\|\Gamma\mathbf{s}_t\|_2^2
=
\mathrm{vec}(\Gamma)^\top
\left(
\sum_{t=1}^{T}\mathbf{s}_t\mathbf{s}_t^{\top}\otimes I_r
\right)
\mathrm{vec}(\Gamma).
\end{equation}
Let
\begin{equation}
\Revision{\mathbf{G}_s}:=\sum_{t=1}^{T}\mathbf{s}_t\mathbf{s}_t^{\top}\in\mathbb{R}^{m\times m}.
\end{equation}
The problem becomes
\begin{equation}
\min_{\mathbf{q}}\;
w_{\mathrm{red}}\|A\mathbf{q}-\mathbf{y}\|_2^2
+
w_u\,\mathbf{q}^{\top}E_{\Gamma}^{\top}(\Revision{\mathbf{G}_s}\otimes I_r)E_{\Gamma}\mathbf{q}
\quad
\text{s.t.}\quad
\Revision{A_{\mathrm{cl}}}\mathbf{q}=0.
\end{equation}

Its KKT system is
\begin{equation}
\begin{bmatrix}
\Revision{Q} & \Revision{A_{\mathrm{cl}}}^{\top}\\
\Revision{A_{\mathrm{cl}}} & 0
\end{bmatrix}
\begin{bmatrix}
\mathbf{q}\\
\boldsymbol{\nu}
\end{bmatrix}
=
\begin{bmatrix}
2w_{\mathrm{red}}A^{\top}\mathbf{y}\\
0
\end{bmatrix},
\end{equation}
with
\begin{equation}
\Revision{Q}=
2\left(
w_{\mathrm{red}}A^{\top}A
+
w_u\,E_{\Gamma}^{\top}(\Revision{\mathbf{G}_s}\otimes I_r)E_{\Gamma}
\right).
\end{equation}

After solving for \((\mathbf{q},\boldsymbol{\nu})\), we recover
\(\tilde{\mathbf{z}}_1\) and \(\Gamma\), then generate the cyclic reduced trajectory by
\begin{equation}
\tilde{\mathbf{z}}_{t+1}=\hat{\mathbf{K}}\tilde{\mathbf{z}}_t+\Gamma\mathbf{s}_t.
\end{equation}

\begin{figure}[]
    \centering
    \includegraphics[width=\linewidth]{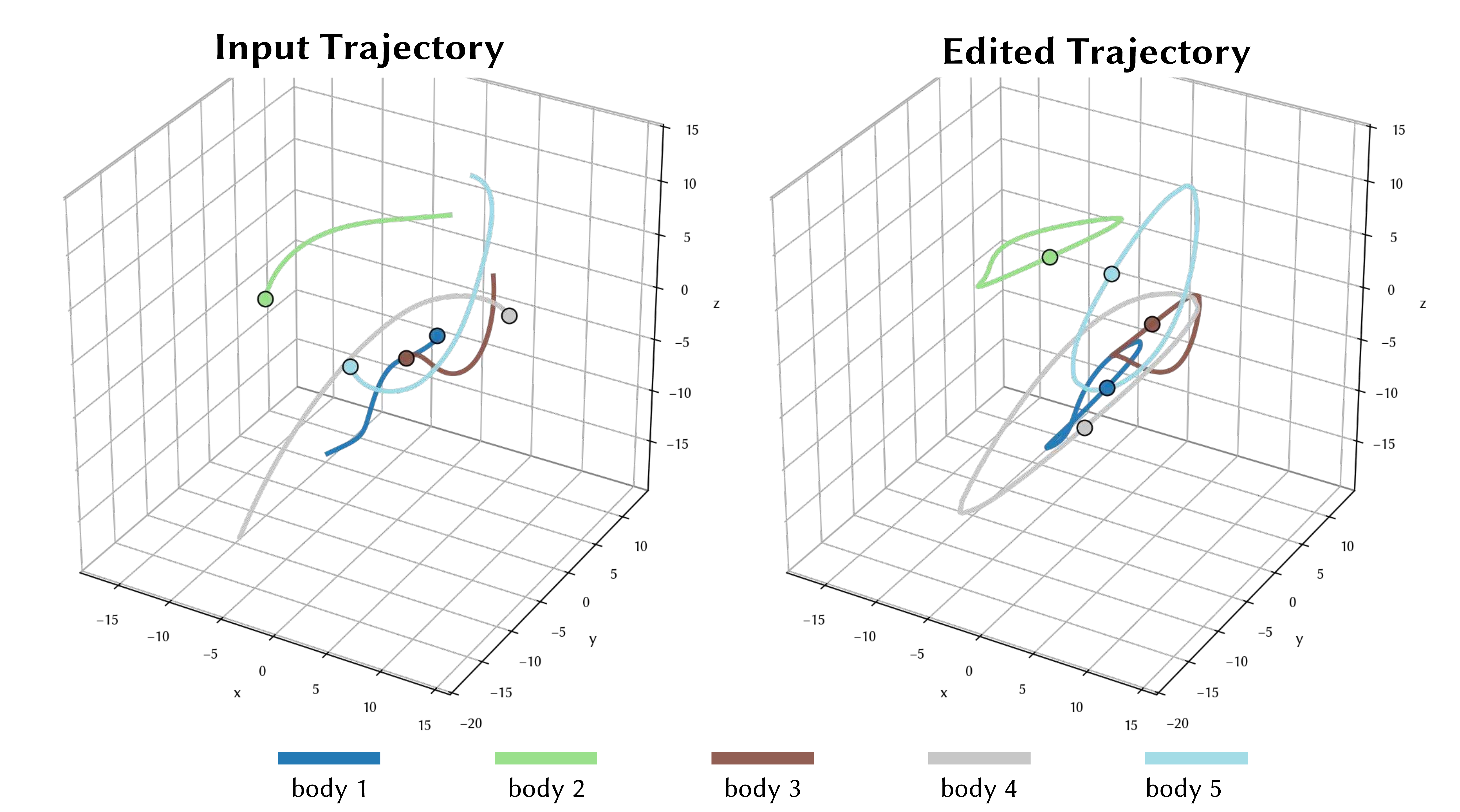}
    \caption{N-body Dynamics. A 5-body system of different mass interacting under mutual gravity. Left: the original Newtonian trajectories are non-cyclic within the given sequence. Right: our method synthesizes a cyclic trajectory with the same period as the input while preserving the overall structure of the given physical trajectory. Each body's path is transformed into a closed, ring-like loop, producing a seamless animation.}
    \label{fig:nbody}
\end{figure}

\section{Results}

We evaluate our method on several representative physical systems: N-body dynamics (Sec.~\ref{se:nbody}), cloth simulation (flag and pinned cloth, Sec.~\ref{se:cloth}), deformable solids (cute dragon and parrot, Sec.~\ref{se:Deformable Solids}), and shallow water (wave propagation and SCA-shaped pattern, Sec.~\ref{se:water}). N-body dynamics data are generated by numerically integrating Newton's equations using the velocity-Verlet scheme, implemented in NumPy \cite{harris2020array}; cloth simulation data are generated by Stiff-GIPC \cite{huang2025stiffgipc}; deformable solids data are generated by stable neo-hookean flesh simulator \cite{smith2018stable}, and shallow water data are generated by a 2D finite-difference solver based on the shallow-water equations of de Saint-Venant (1871), implemented in NumPy \cite{harris2020array}. For all examples, experiments were conducted on a desktop machine equipped with a 12th Gen Intel(R) Core(TM) i7-12700KF CPU. Owing to the equation-free nature of our method, the same pipeline applies across all examples with different underlying dynamics. Across all examples, our method converts non-cyclic input trajectories into seamless loops. Because the cyclicity is imposed as a hard constraint, the synthesized trajectories satisfy the closure condition up to machine precision (on the order of \(10^{-10}\)). In addition, we demonstrate an interactive local-control extension in Sec~\ref{se:interactive}.

For each system, we construct the state vector by stacking the quantities that describe its dynamics: positions and velocities for N-body, cloth, and deformable solids, and water height together with horizontal velocities for shallow water. For a trajectory of length \(T+1\), we fit the surrogate dynamics on the first \(T\) frames and reserve the final frame for evaluating cyclic closure of the synthesized trajectory. Table~\ref{tab:result_summary} summarizes the state dimensions, the numbers of input and fitting frames, the subspace dimensions, the control basis dimensions, and the computation time for all examples.


\begin{table*}[t]
\centering
\small
\setlength{\tabcolsep}{5pt}
\renewcommand{\arraystretch}{1.08}
{\footnotesize \textsuperscript{\dag}Machine precision is on the order of \(10^{-10}\).\par}
\vspace{3pt}
\begin{tabular}{lcccccc}
\toprule
& & & & & \multicolumn{2}{c}{\shortstack[c]{Computation time (s)\\[0pt]\rule{3.2cm}{0.4pt}}}  \vspace{-9pt}\\

Example 
& \shortstack[c]{State\\dim. (\(n\))} 
& \shortstack[c]{Frames\\(input / fit)} 
& \shortstack[c]{Subspace\\dim. (\(r\))} 
& \shortstack[c]{Control basis\\dim. (\(m\))} 
& \shortstack[c]{Processing} 
& \shortstack[c]{Optimization} \\
\midrule
N-body (Fig.~\ref{fig:nbody})                 & \(5 \times 6\)      & 401 / 400 & 3  & 16 & 0.001 & 0.033 \\
Flag (Fig.~\ref{fig:flag})                    & \(4225 \times 6\)   & 101 / 100 & 8  & 16 & 0.479 & 0.015 \\
Pinned cloth (Fig.~\ref{fig:pinned_cloth})    & \(4225 \times 6\)   & 201 / 200 & 8  & 16 & 0.513 & 0.013 \\
Cute dragon (Fig.~\ref{fig:cute_dragon})      & \(4243 \times 6\)   & 301 / 300 & 8  & 16 & 0.798 & 0.019 \\
Parrot (Fig.~\ref{fig:parrot})                & \(10763 \times 6\)  & 101 / 100 & 16 & 16 & 0.558 & 0.025 \\
Wave propagation (Fig.~\ref{fig:shallow_water})  & \(22500 \times 3\)  & 101 / 100 & 16 & 16 & 0.587 & 0.027 \\
SCA-shaped pattern (Fig.~\ref{fig:shallow_water_sca})  & \(131072 \times 3\)  & 201 / 200 & 16 & 16 & 4.581 & 0.044 \\
\bottomrule
\end{tabular}
\caption{Summary of the examples used in our evaluation. We report the full-state dimension, the numbers of input and fitting frames, the reduced space dimension, the control basis dimension, and the computation time, separated into preprocessing time and optimization time. For all examples, we fix \(w_{\mathrm{red}}=10^{-2}\) and \(w_u=3\). The cyclic constraint error is at machine precision\textsuperscript{\dag}.}
\label{tab:result_summary}
\end{table*}

\Revision{
\subsection{Cyclic Editing Across Diverse Systems}

\subsubsection{N-body Simulation}\label{se:nbody}

Newtonian $N$-body systems often exhibit chaotic dynamics, and trajectories are generally not cyclic over a finite time window \cite{LI2019collisionless}. In this example, we take a non-cyclic input sequence and synthesize a cyclic trajectory by enforcing exact temporal closure as a hard constraint. As shown in Fig.~\ref{fig:nbody}, the edited trajectory remains close to the original gravitational motion and preserves the overall orbital configuration, while forming a seamless loop.

\begin{figure}[]
    \centering
    \includegraphics[width=\linewidth]{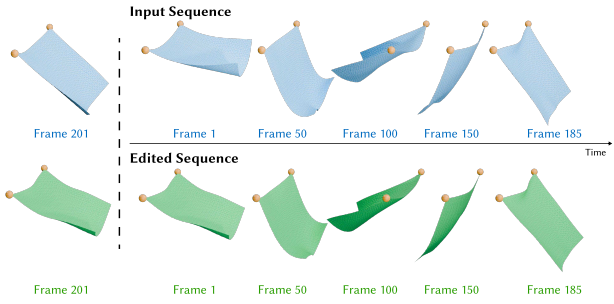}
    \caption{Pinned-cloth. Top: the input sequence is non-cyclic, since the final configuration does not return to the initial state within the given duration. Bottom: our edited sequence becomes cyclic while preserving the dominant deformation behavior of the cloth. The dashed line marks the temporal boundary between the last and first frames, and the edited result closes this boundary smoothly to produce a seamless loop.}
    \label{fig:pinned_cloth}
\end{figure}

\subsubsection{Cloth Dynamics}\label{se:cloth}

\begin{figure*}[t]
    \centering
    \includegraphics[width=0.98\textwidth]{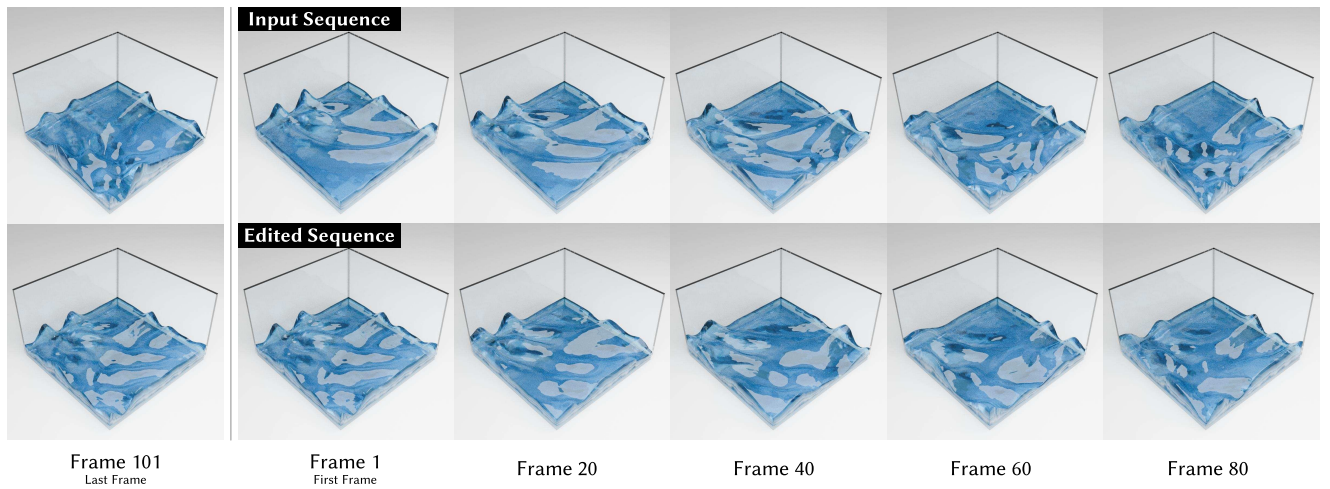}
    \caption{Wave propagation. Top: the input free-surface sequence is non-cyclic, and a visible mismatch appears across the temporal boundary between Frame 101 and Frame 1. Bottom: our edited sequence enforces cyclic closure over the same duration while preserving the dominant wave-front structure and overall surface evolution.}
    \label{fig:shallow_water}
\end{figure*}

We evaluate our method on two challenging cloth examples with distinct dynamical characteristics: a wind-driven flag and a pinned cloth undergoing large-amplitude deformation. Both input sequences exhibit coherent oscillatory motion, but neither is cyclic over the observed time interval, as the final configuration does not return to the initial state.

Fig.~\ref{fig:flag} shows a flag-in-the-wind sequence with visually periodic fluttering. Although the motion contains a dominant oscillatory pattern, the input trajectory does not close in time. Our method synthesizes a cyclic sequence of the same duration under the learned surrogate dynamics, producing a seamless loop while preserving the characteristic fluttering behavior of the flag. In particular, the edited motion maintains the dominant temporal frequency and the spatial wave-like deformation patterns observed in the input.

Fig.~\ref{fig:pinned_cloth} shows a pinned-cloth sequence with larger-amplitude nonlinear deformation. This example is less regular than the flag sequence and involves stronger global shape changes over time. Nevertheless, our method smoothly closes the temporal boundary without directly stitching the endpoints. The resulting cyclic motion remains visually consistent with the input and preserves the dominant deformation modes, while applying only the necessary control to guide the trajectory back to its initial configuration.

Together, these examples demonstrate that our method can synthesize seamless cyclic cloth animations across different regimes of cloth motion, from wind-driven fluttering to large-amplitude nonlinear deformation, without relying on naive endpoint interpolation or abrupt frame matching.

\subsection{Deformable Solids}\label{se:Deformable Solids}


We evaluate our method on two deformable-solid examples that highlight different use cases: cyclic editing of an existing deformable sequence and synthesis of a new cyclic animation from a non-cyclic motion prior.

Fig.~\ref{fig:cute_dragon} shows a deforming dragon sequence. Although the input motion is physically coherent, it does not return to its initial configuration within the observed time window. Our method edits the sequence into a cyclic one over the same duration under the learned surrogate dynamics. The edited result remains visually close to the input while preserving the overall deformation pattern, demonstrating that our method can close a deformable trajectory without introducing an abrupt endpoint correction.

Fig.~\ref{fig:parrot} shows a deformable parrot undergoing wing flapping. In this example, the input provides a characteristic flapping motion, but the full trajectory is not directly usable as a loop due to accumulated global drift. Our method uses this non-cyclic motion as a prior and synthesizes a new cyclic animation with the same duration. The generated motion preserves the recognizable wing-flapping behavior while removing the long-term drift, producing a repeatable animation that transitions smoothly from the final frame back to the first.

Together, these examples show that our method applies beyond endpoint correction of existing deformable sequences. It can also synthesize new cyclic deformable animations from non-cyclic inputs, while preserving the dominant motion characteristics of the original trajectory.

\begin{figure}[t]
    \centering
    \includegraphics[width=\linewidth]{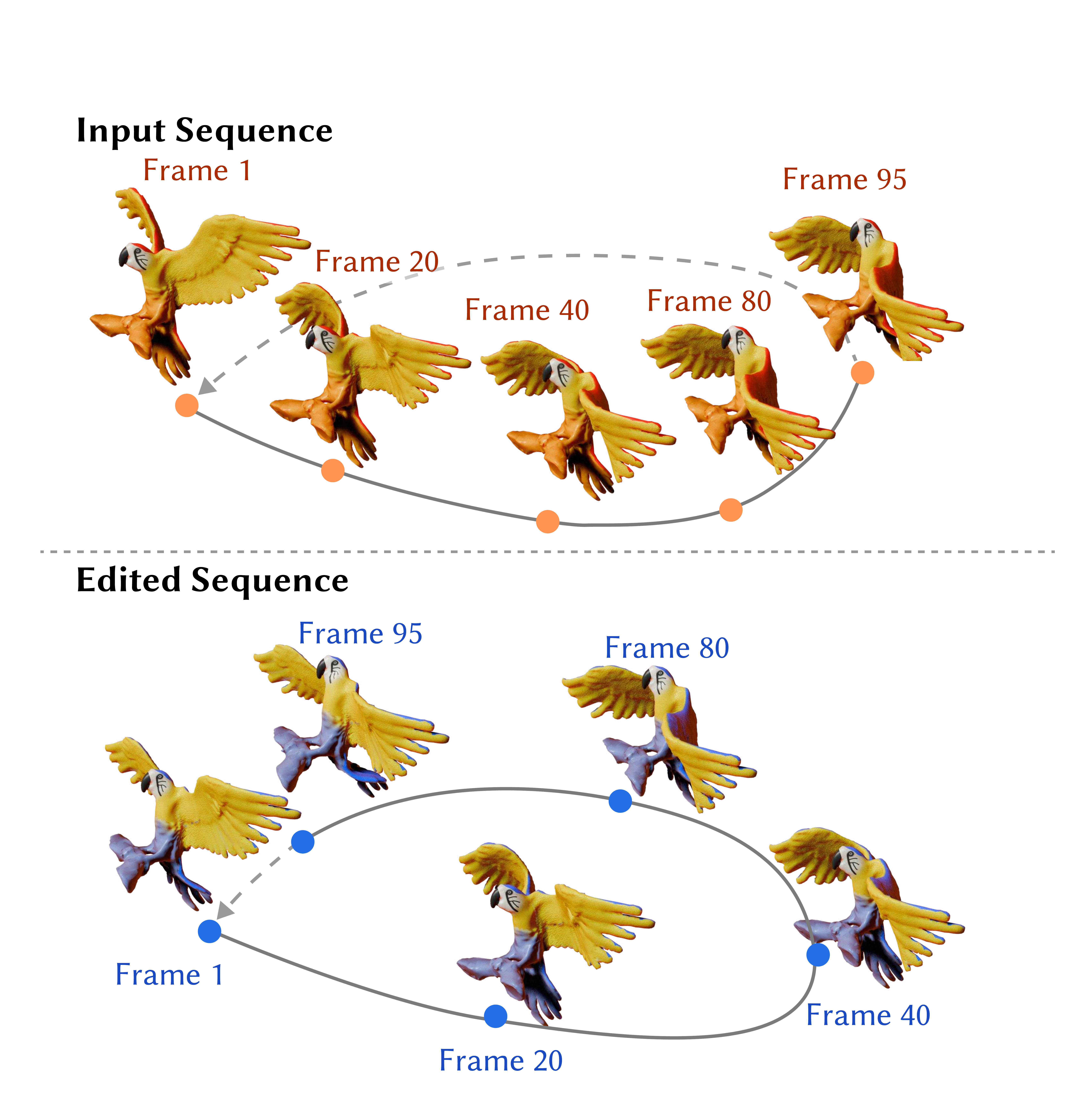}
    \caption{Parrot. Top: the input sequence is non-cyclic, as the wing-flapping movement gradually drifts downward instead of returning to the starting configuration. Bottom: our method edits the sequence into a cyclic one over the same duration. Near the end of the sequence, the downward drift is corrected and the flapping pattern is guided back toward the first frame, producing a seamless loop.}
    \label{fig:parrot}
\end{figure}



\subsubsection{Shallow Water}\label{se:water}
We evaluate our method on two shallow-water sequences that highlight different challenges in cyclic fluid animation: preserving coherent wave propagation and recovering a recognizable structured pattern after dispersion.

Fig.~\ref{fig:shallow_water} shows a shallow-water sequence dominated by propagating wave fronts. Although the free surface evolves coherently over time, the last frame does not return to the first, producing a visible discontinuity under looped playback. Our method synthesizes a cyclic sequence with the same duration under the learned equation-free surrogate dynamics. The edited result preserves the main fluid behaviors of the input, including the overall rise-and-fall motion of the free surface, the dominant crest-and-trough structures, and the primary wave propagation trends.

Fig.~\ref{fig:shallow_water_sca} shows a more structured shallow-water example initialized with a recognizable SCA-shaped free-surface pattern. Compared with the previous sequence, this example contains richer spatial and temporal variation, and the initial pattern gradually disperses over time rather than reappearing at the end of the input trajectory. Our method synthesizes a cyclic animation that restores the SCA-shaped pattern near the temporal boundary while maintaining the overall wave evolution throughout the sequence.

Together, these examples demonstrate that our method can produce seamless cyclic shallow-water animations across both generic wave propagation and structured dispersive patterns, without directly stitching the endpoint frames.




}

\begin{figure}[t]
    \centering
    \includegraphics[width=\linewidth]{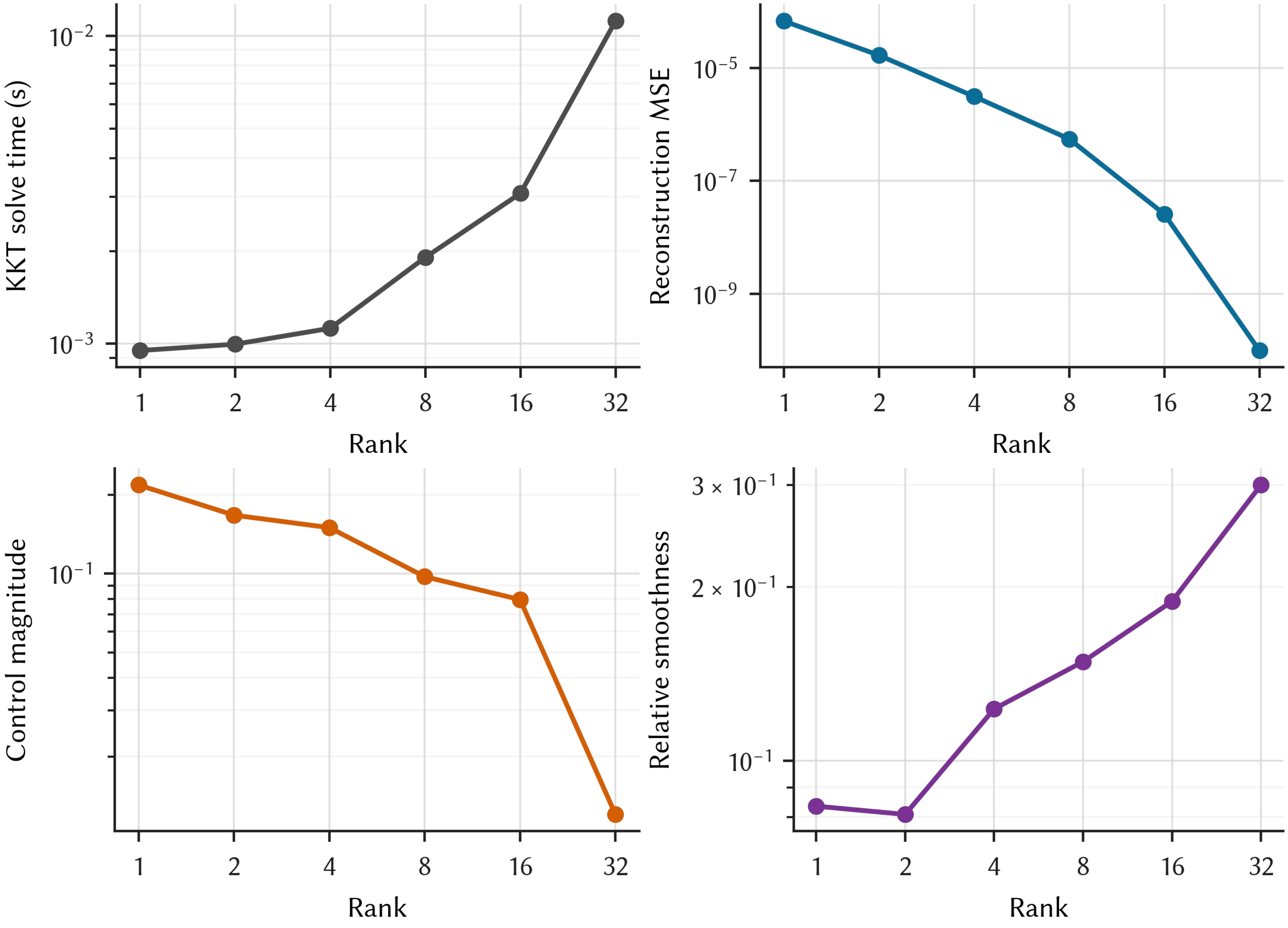}
    \caption{\Revision{Effect of subspace dimension. Increasing the rank improves position reconstruction and reduces the required corrective control, while increasing KKT solve time and relative smoothness. This shows the tradeoff between reconstruction fidelity, control effort, temporal smoothness, and computational cost when choosing the reduced subspace dimension.}}
    \label{fig:parrot_rank}
\end{figure}

\Revision{\subsection{Extension: Interactive Control}\label{se:interactive}

\begin{figure*}[t]
    \centering
    \includegraphics[width=0.98\textwidth]{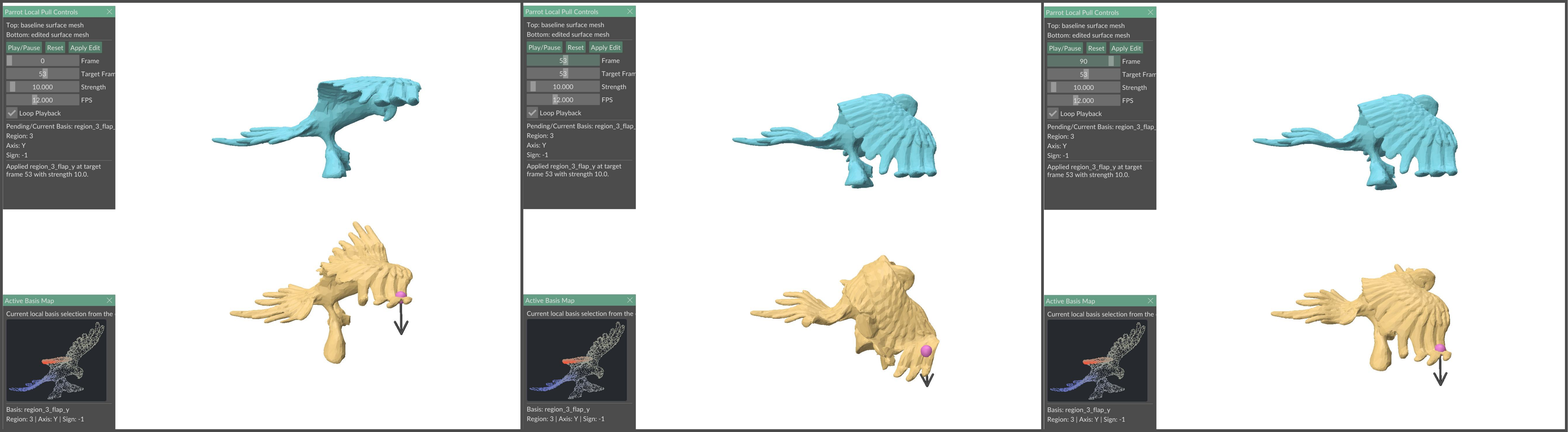}
    \caption{Interactive local control on the parrot sequence. Top: The synthesized cyclic sequence without user control. Bottom: applying a downward local edit at the wing tip (frame 53, strength 10) produces a stronger downward wing motion while preserving cyclic playback.}
    \label{fig:interactive_control}
\end{figure*}

We further extend our formulation to support user-guided local \Revision{control input} while still imposing cyclicity. To let the user impose the \Revision{local control input}, we factorize the coefficient matrix as \(\Gamma = \Revision{B}\Revision{\Lambda}\), where \(\Revision{B}\in\mathbb{R}^{r\times k}\) contains a set of user-defined localized spatial control bases, similar in spirit to prior work on reduced-space control and localized deformation bases \cite{Harmon2013subspace,Benchekroun2025force}, and \(\Revision{\Lambda}\in\mathbb{R}^{k\times m}\) represents the temporal coefficients of bases. 

A user interaction specifies a spatial region, a control direction, and a target frame. The selected region and direction determine an active local basis index \(j^\star\), while a smooth temporal target profile \(a_t\) prescribes when the edit should be most prominent. We then solve for a cyclic sequence that remains close to the input sequence while matching the requested local edit:
\begin{equation}
\begin{aligned}
\min_{\tilde{\mathbf{z}}_1,\,\Revision{\Lambda}}\;&
w_{\mathrm{red}}\sum_{t=1}^{T}
\|\tilde{\mathbf{z}}_t-\mathbf{z}_t^{\mathrm{input}}\|_2^2
+w_{u}\|\Revision{\Lambda}\|_F^2\\
&\quad
+w_{\mathrm{profile}}\sum_{t=1}^{T}\big((\Revision{\Lambda}\mathbf{s}_t)_{j^\star}-a_t\big)^2 \\
\text{s.t.}\;&
\tilde{\mathbf{z}}_{t+1}
=
\hat{\mathbf{K}}\tilde{\mathbf{z}}_t
+
\Revision{B}\Revision{\Lambda}\mathbf{s}_t,
\qquad t=1,\ldots,T,\\
&
\tilde{\mathbf{z}}_{T+1}=\tilde{\mathbf{z}}_1.
\end{aligned}
\label{eq:interactive_control}
\end{equation}

Because the extension remains a low-dimensional linearly constrained quadratic program, each user edit can be updated in sub-second time. Details are provided in Appendix~\ref{sec:appendix_interactive_control}.

Fig.~\ref{fig:interactive_control} shows an example on the parrot sequence.
The user selects a local region near the wing tip and applies a downward edit around frame 53, with the control strength set to 10. The resulting sequence follows the requested local deformation while preserving the overall flapping pattern and cyclicity.

}

\Revision{\subsection{Evaluation}

\subsubsection{Effect of Subspace Dimension}
To evaluate the effect of different subspace dimensions, we vary the rank of the reduced space on the parrot example using ($r=1,2,4,8,16,32$), while keeping all other parameters fixed as reported in Table~\ref{tab:result_summary}. We measure reconstruction MSE as the mean squared error between the input positions and the edited positions reconstructed from the reduced trajectory, average control magnitude as the time-averaged \(L_2\) norm of the control input, and relative smoothness as acceleration normalized by velocity. Since cyclicity is imposed as a hard constraint, all edited sequences are closed by construction.

As shown in Fig.~\ref{fig:parrot_rank}, increasing the rank leads to a larger KKT system, with solve time increasing from $8.46\times10^{-4}\mathrm{s}$ to $1.79\times10^{-2}\mathrm{s}$. At the same time, higher-rank subspaces provide a more expressive representation of the input trajectory: the reconstruction MSE decreases from $6.75\times10^{-5}$ to $9.85\times10^{-11}$, while the average control magnitude drops from $2.17\times10^{-1}$ to $1.20\times10^{-2}$. This indicates that richer subspaces require less corrective control to synthesize a cyclic motion.

The visual differences are also apparent in Fig.~\ref{fig:parrot_rank_figures}. Low-rank subspaces produce smoother motions with reduced spatial and temporal variation. As the rank increases, more high-frequency details from the original trajectory are retained, leading to larger motion amplitudes and a closer reproduction of the input dynamics. The relative smoothness metric increases from $8.32\times10^{-2}$ to $3.01\times10^{-1}$ as the rank grows, indicating that higher-rank spaces preserve more temporal variation while lower-rank spaces act as a stronger smoothing prior. Overall, these results demonstrate a tradeoff between runtime, reconstruction fidelity, control effort, and motion smoothness when selecting the reduced-space dimension.}
\begin{figure}[ht]
    \centering
    \includegraphics[width=1\linewidth]{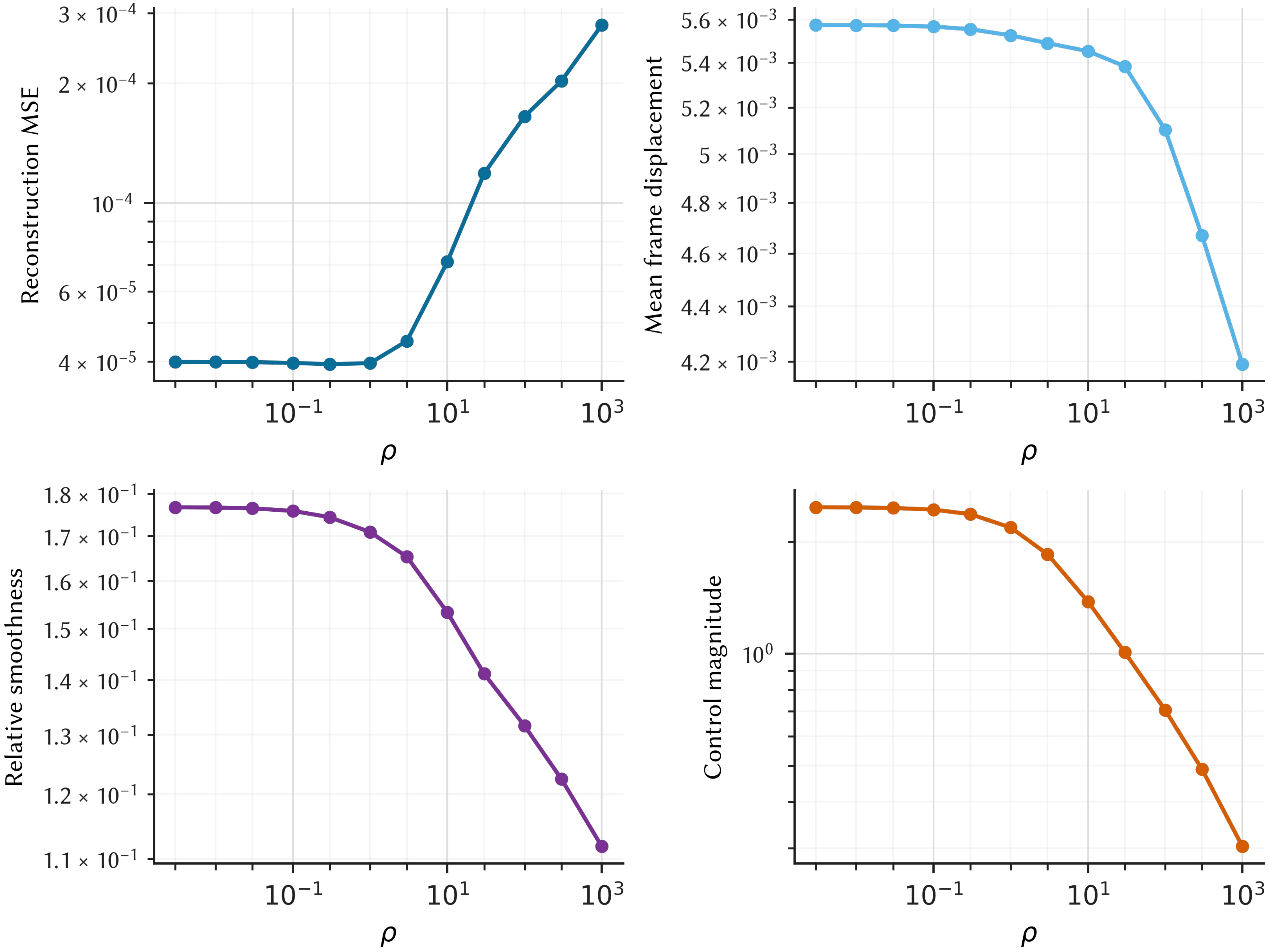}
    \caption{\Revision{Effect of the regularization ratio $\rho = w_u/w_{\mathrm{red}}$, evaluated at ratios $0.01$, $0.1$, $1$, $10$, $100$, $300$, and $1000$. Increasing this ratio penalizes corrective control more strongly, which reduces the average control magnitude but increases reconstruction MSE. Decreasing the ratio places more weight on trajectory fidelity, yielding closer agreement with the input at the cost of stronger corrective control.}}
    \label{fig:regularization_ratio}
\end{figure}

\Revision{\subsubsection{Effect of Regularization Weights}
To study the influence of the regularization weights in Eq.~\eqref{eq:optimization_problem_control_cyclic}, we perform parameter sweeps on the flag example while keeping all other parameters fixed, as reported in Table~\ref{tab:result_summary}. The objective contains a trajectory-fidelity term weighted by $w_{\mathrm{red}}$ and a corrective-control regularization term weighted by $w_u$, and their relative ratio $\rho = w_u/w_{\mathrm{red}}$ controls the fidelity-control tradeoff: a larger ratio penalizes corrective control more strongly, while a smaller ratio emphasizes agreement with the input trajectory. We also report mean frame displacement, which measures the average vertex displacement between consecutive edited frames.

As shown in Fig.~\ref{fig:regularization_ratio}, increasing the ratio $\rho$ makes the optimizer penalize corrective control more strongly: the average control magnitude decreases, while the reconstruction MSE increases because the edited trajectory is allowed to deviate further from the observed sequence. Conversely, a smaller ratio emphasizes trajectory fidelity, reducing reconstruction MSE but requiring stronger corrective control to satisfy the cyclic constraint. These results demonstrate that the ratio $\rho$ provides intuitive control over the balance between trajectory fidelity and the amount of corrective control used to synthesize cyclic animation.}
\begin{figure*}
    \centering
    \includegraphics[width=\textwidth]{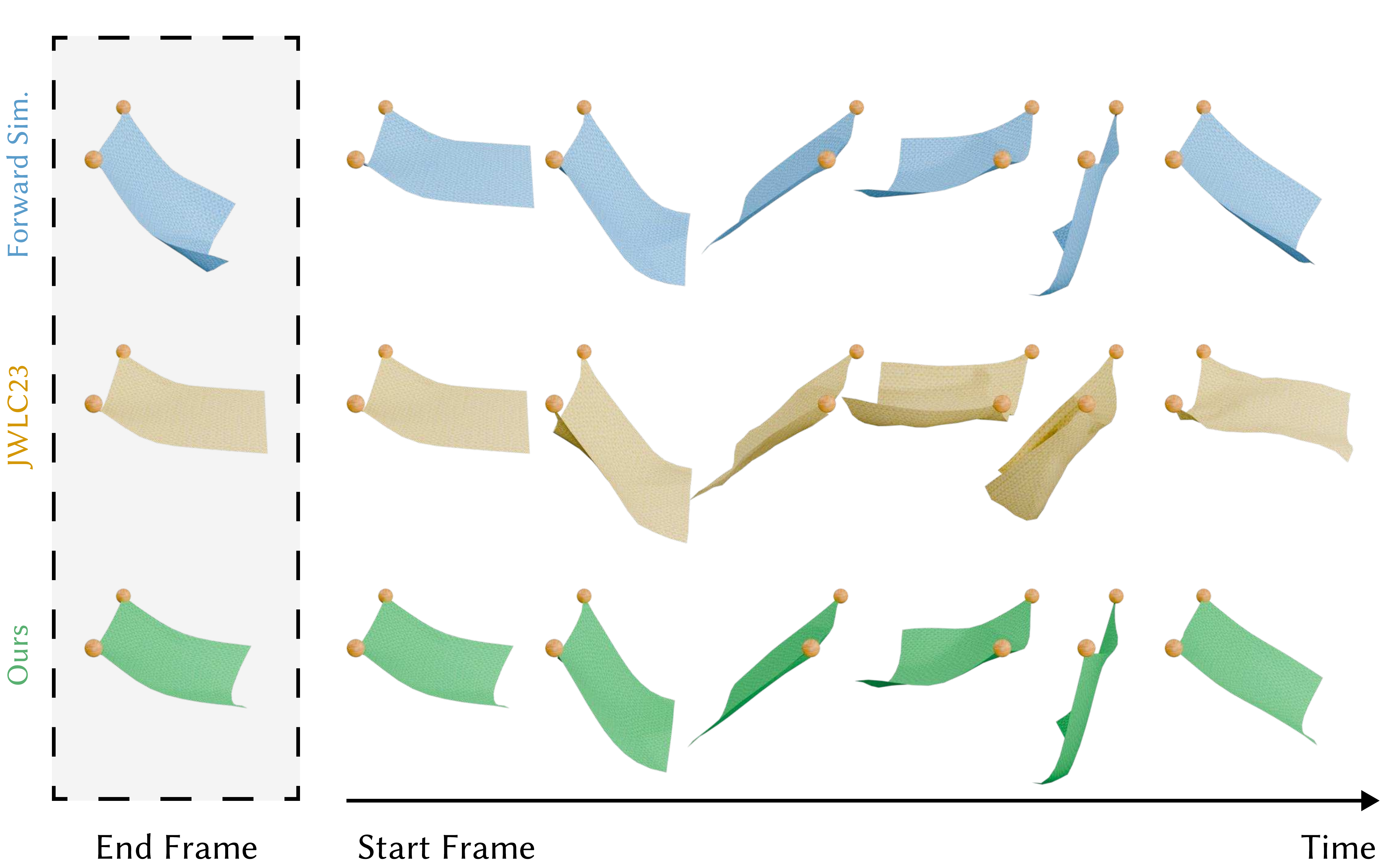}
    \caption{\Revision{Method comparison. The forward simulation does not naturally return to its start state. We compare our method with~\cite{Jia2023physical} under the closest comparable setting: their method generates a cyclic simulation from the first two frames of the forward simulation, while ours uses the full forward trajectory to synthesize a cyclic animation. Although the inputs and assumptions differ, both methods produce looped motions with similar overall deformation behavior.}}
    \label{fig:comparison_with_others}
\end{figure*}

\begin{table}[t]
\centering
\setlength{\tabcolsep}{8pt}
\renewcommand{\arraystretch}{1.08}
\begin{tabular}{>{\small}c|cc}
\toprule
Example & \shortstack[c]{Mean Physical \\ Quantity} & \shortstack[c]{Edited/Input \\ Quantity Ratio} \\
\midrule
Flag (Fig.~\ref{fig:flag}) & Elastic energy & $1.241$\\
Parrot (Fig.~\ref{fig:parrot}) & Deformation energy & $1.340$\\
Shallow water (Fig.~\ref{fig:shallow_water}) & Volume & $1.000$\\
\bottomrule
\end{tabular}
\caption{\Revision{Physical diagnostics on representative examples. Method parameters are the same as in Table~\ref{tab:result_summary}.
For each example, we report the mean (over time) of a physically relevant quantity for the edited trajectory, and its ratio to the corresponding mean of the input trajectory. Ratios close to one indicate that the cyclic edit remains within a similar physical scale as the observed motion.}}
\label{tab:physical_diagnostics}
\end{table}

\begin{table}[t]
\centering
\setlength{\tabcolsep}{8pt}
\renewcommand{\arraystretch}{1.08}
\begin{tabular}{c|cc}
\toprule
\shortstack[c]{Subspace \\ dim.\ ($r$)}
& \shortstack[c]{Edited Mean \\ Elastic Energy}
& \shortstack[c]{Edited/Input \\ Elastic Ratio} \\
\midrule
$1$  & $5.80\times10^{-3}$ & $2.000$\\
$2$  & $5.10\times10^{-3}$ & $1.759$\\
$4$  & $4.01\times10^{-3}$ & $1.383$\\
$8$  & $3.60\times10^{-3}$ & $1.241$\\
$16$ & $3.30\times10^{-3}$ & $1.138$\\
$32$ & $3.03\times10^{-3}$ & $1.045$\\
\bottomrule
\end{tabular}
\caption{\Revision{Mean elastic energy diagnostics on the flag.
We report the mean elastic energy of the edited animation and the edited‑to‑input ratio. The input mean elastic energy is $2.90\times10^{-3}$.}}
\label{tab:energy_flag}
\end{table}
\Revision{\subsubsection{Physical Diagnostics and Robustness}
Because our method observes only trajectories and does not assume access to the underlying simulator or physical model, we do not evaluate physical accuracy against an external reference solution. Instead, we use the input trajectory as a reference and report diagnostic quantities that measure whether the cyclic edit remains within a comparable physical scale. These diagnostics are not enforced by our optimization, but serve as a simple check on the physical plausibility of the edited motion relative to the observed sequence.

We evaluate representative examples using diagnostics matched to their physical behavior. Since each trajectory consists of multiple frames, we report time-averaged quantities over the sequence. For the flag example, we use the mean unit-stiffness edge-spring elastic energy, with rest lengths taken from the rest shape. For the parrot example, we measure the mean deformation-energy that captures changes in the articulated deformation. For the shallow-water example, we measure the mean fluid volume to assess volume preservation. As summarized in Table~\ref{tab:physical_diagnostics}, the edited trajectories remain close to the input scale across these different diagnostics. This suggests that although our method modifies the observed trajectory to enforce cyclicity, it remains within a comparable physical scale as the input.

We further study this behavior under different reduced subspace dimensions on the flag example (Fig.~\ref{fig:flag_rank_figures}). As reported in Table~\ref{tab:energy_flag}, the edited-to-input elastic-energy ratio decreases from $2.000$ at $r=1$ to $1.045$ at $r=32$. This trend is consistent with the reconstruction and control metrics above: very low-dimensional subspaces require larger corrective deformation to satisfy the cyclic constraint, while sufficiently expressive subspaces can achieve cyclicity with only modest changes to the input deformation scale.}

\Revision{\subsection{Relation to Prior Work}
The physical cyclic animation method of~\cite{Jia2023physical} is closely related to our goal of producing cyclic motion. The two methods, however, differ in scope and assumptions. Their formulation is physics-based and model-dependent, requiring access to the underlying governing equations and a physical simulator for optimization. In contrast, our method is equation-free and trajectory-only. All information used by our method comes from the observed motion itself.

This difference makes a direct comparison inherently imperfect. Nevertheless, we construct the closest comparable experiment we can. We generate a forward pinned-cloth trajectory using C-IPC~\cite{li2021cipc} and use this trajectory as the common reference. For~\cite{Jia2023physical}, we follow their proposed setting and optimization procedure, which uses the first two frames of the forward simulation as initialization conditions to generate a cyclic animation with the same period. For our method, we use the forward simulation frames as the observed trajectory and synthesize a cyclic edit directly from these frames, without using the simulator or physical model.

As shown in Fig.~\ref{fig:comparison_with_others}, the forward simulation does not naturally return to its start state, whereas both methods produce cyclic motions. Although the inputs and assumptions are different, the resulting animations exhibit similar overall deformation behavior. We also compare the mean elastic energy relative to the forward simulation in Table~\ref{tab:prior_energy_comparison}. The method of~\cite{Jia2023physical} has an average elastic-energy ratio of $3.385$, while our result remains closer to the input energy scale with a ratio of $1.390$.}

\begin{table}[t]
\centering
\setlength{\tabcolsep}{8pt}
\renewcommand{\arraystretch}{1.08}
\begin{tabular}{c|cc}
\toprule
Method
& \shortstack[c]{Mean Elastic \\ Energy}
& \shortstack[c]{Edited/Input \\ Elastic Ratio} \\
\midrule
Forward Sim. & $6.784\times10^{-3}$ & $1.000$\\
\cite{Jia2023physical} & $2.297\times10^{-2}$ & $3.385$\\
Ours & $9.430\times10^{-3}$ & $1.390$\\
\bottomrule
\end{tabular}
\caption{\Revision{Mean elastic energy diagnostics on the pinned-cloth comparison.
We use the same unit-stiffness edge-spring energy as above and report the input-relative elastic ratio. The forward simulation has a mean elastic energy of $6.784\times10^{-3}$.}}
\label{tab:prior_energy_comparison}
\end{table}

\Revision{\section{Discussion and Limitations}
Our method works directly from a single observed trajectory and does not attempt to model the underlying physical system.  Consequently, it relies solely on the information contained in that trajectory and requires no governing equations or other system-specific physical data.

Within this formulation and given data, cyclicity and dynamical consistency are expressed through the learned operator and the optimization objective. The generated trajectory is constrained to form a closed cycle while remaining close to the observed behavior and requiring only limited corrective control. However, this notion of dynamical consistency is defined with respect to the observed trajectory dynamics, rather than a fully specified physical system. Because the method observes only a trajectory, it can approximate the temporal evolution present in the input data, but it cannot recover the complete governing dynamics of the system. Therefore, our method cannot guarantee exact physical consistency with the unknown true system; instead, it seeks a cyclic edit that is \textbf{visually plausible and temporally coherent relative to the observed trajectory}.}

\section{Conclusion}
We present a \Revision{trajectory-editing pipeline} that converts non-cyclic animations into cyclic ones while \Revision{preserving the characteristic dynamics observed in the input}. Our approach learns a Koopman surrogate from the input sequence and enforces periodicity through a minimal \Revision{control input} under a hard temporal constraint. The linear structure of the surrogate leads to an efficient linear solve for cyclic reconstruction, while remaining agnostic to the governing equations of the input data. We demonstrate the generality of our method across a range of simulations, including cloth, deformable solids, and fluids, \Revision{producing cyclic edits that achieve exact temporal closure while retaining the dominant temporal structure and overall character of the observed motion}.

Looking forward, an important direction is extending our framework to more complex settings involving coupled systems, such as fluid–solid interactions, where multiple physical processes interact across different time scales. Another promising avenue is to expand our editing capabilities to support richer and more localized control, enabling users to specify more complex temporal and spatial constraints while maintaining dynamic consistency.

\bibliographystyle{eg-alpha-doi}
\bibliography{bib/refs}

@article{Proctor2018generalizing,
author = {Proctor, Joshua L. and Brunton, Steven L. and Kutz, J. Nathan},
title = {Generalizing Koopman Theory to Allow for Inputs and Control},
journal = {SIAM Journal on Applied Dynamical Systems},
volume = {17},
number = {1},
pages = {909-930},
year = {2018},
doi = {10.1137/16M1062296},
URL = {https://doi.org/10.1137/16M1062296},
eprint = {https://doi.org/10.1137/16M1062296}
}

@InProceedings{Li_2024_CVPR,
    author    = {Li, Zhengqi and Tucker, Richard and Snavely, Noah and Holynski, Aleksander},
    title     = {Generative Image Dynamics},
    booktitle = {Proceedings of the IEEE/CVF Conference on Computer Vision and Pattern Recognition (CVPR)},
    month     = {June},
    year      = {2024},
    pages     = {24142-24153}
}

@article{Modi:2024:Simplicits,
author = {Modi, Vismay and Sharp, Nicholas and Perel, Or and Sueda, Shinjiro and Levin, David I. W.},
title = {Simplicits: Mesh-Free, Geometry-Agnostic Elastic Simulation},
year = {2024},
issue_date = {July 2024},
publisher = {Association for Computing Machinery},
address = {New York, NY, USA},
volume = {43},
number = {4},
issn = {0730-0301},
url = {https://doi.org/10.1145/3658184},
doi = {10.1145/3658184},
journal = {ACM Trans. Graph.},
month = {jul},
articleno = {117},
numpages = {11}
}

@article{Proctor2016dynamic,
author = {Proctor, Joshua L. and Brunton, Steven L. and Kutz, J. Nathan},
title = {Dynamic Mode Decomposition with Control},
journal = {SIAM Journal on Applied Dynamical Systems},
volume = {15},
number = {1},
pages = {142-161},
year = {2016},
doi = {10.1137/15M1013857},
URL = {https://doi.org/10.1137/15M1013857},
eprint = {https://doi.org/10.1137/15M1013857}
}

@article{Jia2023physical,
author = {Jia, Shiyang and Wang, Stephanie and Li, Tzu-Mao and Chern, Albert},
title = {Physical Cyclic Animations},
year = {2023},
issue_date = {August 2023},
publisher = {Association for Computing Machinery},
address = {New York, NY, USA},
volume = {6},
number = {3},
url = {https://doi.org/10.1145/3606938},
doi = {10.1145/3606938},
journal = {Proc. ACM Comput. Graph. Interact. Tech.},
month = aug,
articleno = {45},
numpages = {18}
}

@inproceedings{Barbic2009control,
author = {Barbi\v{c}, Jernej and da Silva, Marco and Popovi\'{c}, Jovan},
title = {Deformable object animation using reduced optimal control},
year = {2009},
isbn = {9781605587264},
publisher = {Association for Computing Machinery},
address = {New York, NY, USA},
url = {https://doi.org/10.1145/1576246.1531359},
doi = {10.1145/1576246.1531359},
booktitle = {ACM SIGGRAPH 2009 Papers},
articleno = {53},
numpages = {9},
location = {New Orleans, Louisiana},
series = {SIGGRAPH '09}
}

@article{Witkin1988spacetime,
author = {Witkin, Andrew and Kass, Michael},
title = {Spacetime constraints},
year = {1988},
issue_date = {Aug. 1988},
publisher = {Association for Computing Machinery},
address = {New York, NY, USA},
volume = {22},
number = {4},
issn = {0097-8930},
url = {https://doi.org/10.1145/378456.378507},
doi = {10.1145/378456.378507},
journal = {SIGGRAPH Comput. Graph.},
month = jun,
pages = {159–168},
numpages = {10}
}

@inproceedings{Heck2007motion,
author = {Heck, Rachel and Gleicher, Michael},
title = {Parametric motion graphs},
year = {2007},
isbn = {9781595936288},
publisher = {Association for Computing Machinery},
address = {New York, NY, USA},
url = {https://doi.org/10.1145/1230100.1230123},
doi = {10.1145/1230100.1230123},
booktitle = {Proceedings of the 2007 Symposium on Interactive 3D Graphics and Games},
pages = {129–136},
numpages = {8},
location = {Seattle, Washington},
series = {I3D '07}
}

@article{Kovar2002motion,
author = {Kovar, Lucas and Gleicher, Michael and Pighin, Fr\'{e}d\'{e}ric},
title = {Motion graphs},
year = {2002},
issue_date = {July 2002},
publisher = {Association for Computing Machinery},
address = {New York, NY, USA},
volume = {21},
number = {3},
issn = {0730-0301},
url = {https://doi.org/10.1145/566654.566605},
doi = {10.1145/566654.566605},
journal = {ACM Trans. Graph.},
month = jul,
pages = {473–482},
numpages = {10}
}

@article{schmid2010dynamic,
  title={Dynamic mode decomposition of numerical and experimental data},
  author={Schmid, Peter J},
  journal={Journal of fluid mechanics},
  volume={656},
  pages={5--28},
  year={2010},
  publisher={Cambridge University Press}
}

@article{chen2025dmd,
  title = {Fast Subspace Fluid Simulation with a Temporally-Aware Basis},
  author = {Siyuan Chen and Yixin Chen and Jonathan Panuelos and Otman Benchekroun and Yue Chang and Eitan Grinspun and Zhecheng Wang},
  year = {2025},
  journal = {ACM Transactions on Graphics},
}

@article{deAguiar2010Stable,
  author    = {Edilson de Aguiar and Leonid Sigal and Adrien Treuille and Jessica K. Hodgins},
  title     = {Stable Spaces for Real-Time Clothing},
  journal   = {ACM Trans. Graph. (SIGGRAPH)},
  volume    = {29},
  number    = {4},
  articleno = {106},
  year      = {2010},
  doi       = {10.1145/1778765.1778843}
}

@article{chang2026lowrank,
    title        = {Low-Rank Koopman Deformables with Log-Linear Time Integration},
    author       = {Yue Chang and others},
    year         = 2026,
    eprint       = {2602.07687},
    archiveprefix = {arXiv},
    primaryclass = {cs.GR},
    note         = {arXiv:2602.07687},
    url          = {https://arxiv.org/abs/2602.07687},
    doi          = {10.48550/arXiv.2602.07687}
}

@article{starke2022deepphase,
    title        = {DeepPhase: Periodic Autoencoders for Learning Motion Phase Manifolds},
    author       = {Sebastian Starke and Ian Mason and Taku Komura},
    year         = 2022,
    journal      = {ACM Transactions on Graphics},
    volume       = 41,
    number       = 4,
    pages        = {1--13},
    doi          = {10.1145/3528223.3530178},
    url          = {https://doi.org/10.1145/3528223.3530178},
    publisher    = {Association for Computing Machinery}
}

@article{li2022ganimator,
    title        = {GANimator: Neural Motion Synthesis from a Single Sequence},
    author       = {Peizhuo Li and Kfir Aberman and Zihan Zhang and Rana Hanocka and Olga Sorkine-Hornung},
    year         = 2022,
    journal      = {ACM Transactions on Graphics},
    volume       = 41,
    number       = 4,
    pages        = {1--12},
    doi          = {10.1145/3528223.3530157},
    url          = {https://doi.org/10.1145/3528223.3530157},
    publisher    = {Association for Computing Machinery}
}

@article{tedla2025generating,
    title        = {Generating the Past, Present and Future from a Motion-Blurred Image},
    author       = {SaiKiran Tedla and Kelly Zhu and Trevor Canham and Felix Taubner and Michael S. Brown and Kiriakos N. Kutulakos and David B. Lindell},
    year         = 2025,
    journal      = {ACM Transactions on Graphics},
    volume       = 44,
    number       = 6,
    doi          = {10.1145/3763306},
    url          = {https://doi.org/10.1145/3763306},
    publisher    = {Association for Computing Machinery},
    note         = {Presented at SIGGRAPH Asia 2025}
}

@inproceedings{abdal2025dynamic,
    title        = {Dynamic Concepts Personalization from Single Videos},
    author       = {Rameen Abdal and Or Patashnik and Ivan Skorokhodov and Willi Menapace and Aliaksandr Siarohin and Sergey Tulyakov and Daniel Cohen-Or and Kfir Aberman},
    year         = 2025,
    month        = aug,
    booktitle    = {Proceedings of the Special Interest Group on Computer Graphics and Interactive Techniques Conference (SIGGRAPH '25)},
    publisher    = {Association for Computing Machinery},
    address      = {New York, NY, USA},
    doi          = {10.1145/3721238.3730829},
    isbn         = {9798400715402},
    url          = {https://doi.org/10.1145/3721238.3730829},
}

@inproceedings{kansy2025reenact,
    title        = {Reenact Anything: Semantic Video Motion Transfer Using Motion-Textual Inversion},
    author       = {Manuel Kansy and Jacek Naruniec and Christopher Schroers and Markus Gross and Romann M. Weber},
    year         = 2025,
    month        = aug,
    booktitle    = {Proceedings of the Special Interest Group on Computer Graphics and Interactive Techniques Conference (SIGGRAPH '25)},
    publisher    = {Association for Computing Machinery},
    address      = {New York, NY, USA},
    pages        = {156:1--156:12},
    doi          = {10.1145/3721238.3730826},
    isbn         = {9798400715402},
    url          = {https://doi.org/10.1145/3721238.3730826},
}

@article{Barbic2012interactive,
author = {Barbi\v{c}, Jernej and Sin, Funshing and Grinspun, Eitan},
title = {Interactive editing of deformable simulations},
year = {2012},
issue_date = {July 2012},
publisher = {Association for Computing Machinery},
address = {New York, NY, USA},
volume = {31},
number = {4},
issn = {0730-0301},
url = {https://doi.org/10.1145/2185520.2185566},
doi = {10.1145/2185520.2185566},
journal = {ACM Trans. Graph.},
month = jul,
articleno = {70},
numpages = {8}
}

@article{koopman1931hamiltonian,
author = {B. O. Koopman },
title = {Hamiltonian Systems and Transformation in Hilbert Space},
journal = {Proceedings of the National Academy of Sciences},
volume = {17},
number = {5},
pages = {315-318},
year = {1931},
doi = {10.1073/pnas.17.5.315},
URL = {https://www.pnas.org/doi/abs/10.1073/pnas.17.5.315},
eprint = {https://www.pnas.org/doi/pdf/10.1073/pnas.17.5.315}}

@inproceedings{Park2002online,
author = {Park, Sang Il and Shin, Hyun Joon and Shin, Sung Yong},
title = {On-line locomotion generation based on motion blending},
year = {2002},
isbn = {1581135734},
publisher = {Association for Computing Machinery},
address = {New York, NY, USA},
url = {https://doi.org/10.1145/545261.545279},
doi = {10.1145/545261.545279},
booktitle = {Proceedings of the 2002 ACM SIGGRAPH/Eurographics Symposium on Computer Animation},
pages = {105–111},
numpages = {7},
location = {San Antonio, Texas},
series = {SCA '02}
}

@article{Mizuguchi2001data,
author = {Mizuguchi, M. and Buchanan, J. and Calvert, T.},
year = {2001},
month = {01},
pages = {},
title = {Data driven motion transitions for interactive games}
}

@inproceedings{de2010stable,
author = {de Aguiar, Edilson and Sigal, Leonid and Treuille, Adrien and Hodgins, Jessica K.},
title = {Stable spaces for real-time clothing},
year = {2010},
isbn = {9781450302104},
publisher = {Association for Computing Machinery},
address = {New York, NY, USA},
url = {https://doi.org/10.1145/1833349.1778843},
doi = {10.1145/1833349.1778843},
booktitle = {ACM SIGGRAPH 2010 Papers},
articleno = {106},
numpages = {9},
location = {Los Angeles, California},
series = {SIGGRAPH '10}
}

@inproceedings{
chen2023crom,
title={{CROM}: Continuous Reduced-Order Modeling of {PDE}s Using Implicit Neural Representations},
author={Peter Yichen Chen and Jinxu Xiang and Dong Heon Cho and Yue Chang and G A Pershing and Henrique Teles Maia and Maurizio M Chiaramonte and Kevin Thomas Carlberg and Eitan Grinspun},
booktitle={The Eleventh International Conference on Learning Representations },
year={2023},
url={https://openreview.net/forum?id=FUORz1tG8Og}
}

@inproceedings{Chang2023liCROM,
author = {Chang, Yue and Chen, Peter Yichen and Wang, Zhecheng and Chiaramonte, Maurizio M. and Carlberg, Kevin and Grinspun, Eitan},
title = {LiCROM: Linear-Subspace Continuous Reduced Order Modeling with Neural Fields},
year = {2023},
isbn = {9798400703157},
publisher = {Association for Computing Machinery},
address = {New York, NY, USA},
url = {https://doi.org/10.1145/3610548.3618158},
doi = {10.1145/3610548.3618158},
booktitle = {SIGGRAPH Asia 2023 Conference Papers},
articleno = {111},
numpages = {12},
location = {Sydney, NSW, Australia},
series = {SA '23}
}

@article{Chang2025shape,
author = {Chang, Yue and Benchekroun, Otman and Chiaramonte, Maurizio M. and Chen, Peter Yichen and Grinspun, Eitan},
title = {Shape Space Spectra},
year = {2025},
issue_date = {August 2025},
publisher = {Association for Computing Machinery},
address = {New York, NY, USA},
volume = {44},
number = {4},
issn = {0730-0301},
url = {https://doi.org/10.1145/3731148},
doi = {10.1145/3731148},
journal = {ACM Trans. Graph.},
month = jul,
articleno = {121},
numpages = {16}
}

@article{Harmon2013subspace,
author = {Harmon, David and Zorin, Denis},
title = {Subspace integration with local deformations},
year = {2013},
issue_date = {July 2013},
publisher = {Association for Computing Machinery},
address = {New York, NY, USA},
volume = {32},
number = {4},
issn = {0730-0301},
url = {https://doi.org/10.1145/2461912.2461922},
doi = {10.1145/2461912.2461922},
journal = {ACM Trans. Graph.},
month = jul,
articleno = {107},
numpages = {10}
}

@article{Lan2024efficient,
author = {Lan, Lei and Lu, Zixuan and Long, Jingyi and Yuan, Chun and Li, Xuan and He, Xiaowei and Wang, Huamin and Jiang, Chenfanfu and Yang, Yin},
title = {Efficient GPU Cloth Simulation with Non-distance Barriers and Subspace Reuse},
year = {2024},
issue_date = {December 2024},
publisher = {Association for Computing Machinery},
address = {New York, NY, USA},
volume = {43},
number = {6},
issn = {0730-0301},
url = {https://doi.org/10.1145/3687760},
doi = {10.1145/3687760},
journal = {ACM Trans. Graph.},
month = nov,
articleno = {226},
numpages = {16}
}

@inproceedings{Chen2024fluid,
author = {Chen, Yixin and Levin, David and Langlois, Timothy},
title = {Fluid Control with Laplacian Eigenfunctions},
year = {2024},
isbn = {9798400705250},
publisher = {Association for Computing Machinery},
address = {New York, NY, USA},
url = {https://doi.org/10.1145/3641519.3657468},
doi = {10.1145/3641519.3657468},
booktitle = {ACM SIGGRAPH 2024 Conference Papers},
articleno = {44},
numpages = {11},
location = {Denver, CO, USA},
series = {SIGGRAPH '24}
}

@article{Kim2009skipping,
author = {Kim, Theodore and James, Doug L.},
title = {Skipping steps in deformable simulation with online model reduction},
year = {2009},
issue_date = {December 2009},
publisher = {Association for Computing Machinery},
address = {New York, NY, USA},
volume = {28},
number = {5},
issn = {0730-0301},
url = {https://doi.org/10.1145/1618452.1618469},
doi = {10.1145/1618452.1618469},
journal = {ACM Trans. Graph.},
month = dec,
pages = {1–9},
numpages = {9}
}

@article{Neumann2013sparse,
author = {Neumann, Thomas and Varanasi, Kiran and Wenger, Stephan and Wacker, Markus and Magnor, Marcus and Theobalt, Christian},
title = {Sparse localized deformation components},
year = {2013},
issue_date = {November 2013},
publisher = {Association for Computing Machinery},
address = {New York, NY, USA},
volume = {32},
number = {6},
issn = {0730-0301},
url = {https://doi.org/10.1145/2508363.2508417},
doi = {10.1145/2508363.2508417},
journal = {ACM Trans. Graph.},
month = nov,
articleno = {179},
numpages = {10}
}

@article{Benchekroun2025force,
author = {Benchekroun, Otman and Grinspun, Eitan and Chiaramonte, Maurizio and Etter, Philip Allen},
title = {Force-Dual Modes: Subspace Design from Stochastic Forces},
year = {2025},
issue_date = {December 2025},
publisher = {Association for Computing Machinery},
address = {New York, NY, USA},
volume = {44},
number = {6},
issn = {0730-0301},
url = {https://doi.org/10.1145/3763310},
doi = {10.1145/3763310},
journal = {ACM Trans. Graph.},
month = dec,
articleno = {200},
numpages = {14}
}

@article{LI2019collisionless,
title = {Collisionless periodic orbits in the free-fall three-body problem},
journal = {New Astronomy},
volume = {70},
pages = {22-26},
year = {2019},
issn = {1384-1076},
doi = {https://doi.org/10.1016/j.newast.2019.01.003},
url = {https://www.sciencedirect.com/science/article/pii/S1384107618302458},
author = {Xiaoming Li and Shijun Liao}
}

@article{barbic2005realtime,
  author  = {Jernej Barbi{\v{c}} and Doug L. James},
  title   = {Real-Time Subspace Integration for St. Venant-Kirchhoff Deformable Models},
  journal = {ACM Transactions on Graphics (SIGGRAPH)},
  year    = {2005},
  volume  = {24},
  number  = {3},
  pages   = {982--990},
}

@article{smith2018stable,
author = {Smith, Breannan and Goes, Fernando De and Kim, Theodore},
title = {Stable Neo-Hookean Flesh Simulation},
year = {2018},
issue_date = {April 2018},
publisher = {Association for Computing Machinery},
address = {New York, NY, USA},
volume = {37},
number = {2},
issn = {0730-0301},
url = {https://doi.org/10.1145/3180491},
doi = {10.1145/3180491},
journal = {ACM Trans. Graph.},
month = mar,
articleno = {12},
numpages = {15}
}

@Article{harris2020array,
 title         = {Array programming with {NumPy}},
 author        = {Charles R. Harris and K. Jarrod Millman and St{\'{e}}fan J.
                 van der Walt and Ralf Gommers and Pauli Virtanen and David
                 Cournapeau and Eric Wieser and Julian Taylor and Sebastian
                 Berg and Nathaniel J. Smith and Robert Kern and Matti Picus
                 and Stephan Hoyer and Marten H. van Kerkwijk and Matthew
                 Brett and Allan Haldane and Jaime Fern{\'{a}}ndez del
                 R{\'{i}}o and Mark Wiebe and Pearu Peterson and Pierre
                 G{\'{e}}rard-Marchant and Kevin Sheppard and Tyler Reddy and
                 Warren Weckesser and Hameer Abbasi and Christoph Gohlke and
                 Travis E. Oliphant},
 year          = {2020},
 month         = sep,
 journal       = {Nature},
 volume        = {585},
 number        = {7825},
 pages         = {357--362},
 doi           = {10.1038/s41586-020-2649-2},
 publisher     = {Springer Science and Business Media {LLC}},
 url           = {https://doi.org/10.1038/s41586-020-2649-2}
}

@article{huang2025stiffgipc,
      author = {Huang, Kemeng and Lu, Xinyu and Lin, Huancheng and Komura, Taku and Li, Minchen},
      title = {StiffGIPC: Advancing GPU IPC for Stiff Affine-Deformable Simulation},
      year = {2025},
      publisher = {Association for Computing Machinery},
      volume = {44},
      number = {3},
      issn = {0730-0301},
      doi = {10.1145/3735126},
      journal = {ACM Trans. Graph.},
      month = may,
      articleno = {31},
      numpages = {20}
}

@article{li2021cipc,
    author = {Minchen Li and Danny M. Kaufman and Chenfanfu Jiang},
    title = {Codimensional Incremental Potential Contact},
    journal = {ACM Trans. Graph. (SIGGRAPH)},
    year = {2021},
    volume = {40},
    number = {4},
    articleno = {170}
}

\appendix
\clearpage
\section{Details For Interactive Control}
\label{sec:appendix_interactive_control}

This appendix gives the formulation of the interactive control.

Let $\tilde{\mathbf z}_t\in\mathbb{R}^r$ be the reduced state, $\hat{\mathbf K}\in\mathbb{R}^{r\times r}$ the
latent dynamics matrix, $\Revision{B}\in\mathbb{R}^{r\times k}$ the control basis matrix,
and $\mathbf s_t\in\mathbb{R}^m$ the Fourier basis at frame $t$. The edited rollout is
\begin{equation}
\tilde{\mathbf z}_{t+1}=\hat{\mathbf K}\tilde{\mathbf z}_t+\Revision{B}\Revision{\Lambda}\mathbf s_t,\quad t=1,\dots,T,
\end{equation}
with $\Revision{\Lambda}\in\mathbb{R}^{k\times m}$ unknown.
Define
\begin{equation}
\mathbf{q}=\begin{bmatrix}\tilde{\mathbf z}_1\\ \mathrm{vec}(\Revision{\Lambda})\end{bmatrix}\in\mathbb{R}^{r+km}.
\end{equation}

Define $A_t\in\mathbb{R}^{r\times(r+km)}$ such that $\tilde{\mathbf z}_t=A_t \mathbf{q}$.
Initialize
\begin{equation}
A_1=\begin{bmatrix}I_r & 0\end{bmatrix}.
\end{equation}
For each $t$, define
\begin{equation}
U_t=\begin{bmatrix}0 & \Revision{B}\left(\mathbf s_t^\top\otimes I_k\right)\end{bmatrix},
\end{equation}
and recurse
\begin{equation}
A_{t+1}=\hat{\mathbf K}A_t+U_t.
\end{equation}
Then $\tilde{\mathbf z}_t=A_t \mathbf{q}$ for all frames.

\paragraph{Stacked fitting and cyclic closure.}
Stack all frames:
\begin{equation}
\Revision{A_{\mathrm{loc}}}=\begin{bmatrix}A_1\\A_2\\\vdots\\A_T\end{bmatrix},\quad
\mathbf{y}=
\begin{bmatrix}
\mathbf z_1^{\mathrm{input}}\\
\vdots\\
\mathbf z_T^{\mathrm{input}}
\end{bmatrix}.
\end{equation}
Hard cyclic closure is
\begin{equation}
\tilde{\mathbf z}_{T+1}-\tilde{\mathbf z}_1=0
\;\Longleftrightarrow\;
\Revision{A_{\mathrm{cl}}}\mathbf{q}=0,
\end{equation}
with
\begin{equation}
\Revision{A_{\mathrm{cl}}}=(\hat{\mathbf K}A_T+U_T)-A_1.
\end{equation}

\paragraph{Objective terms}.
Let
\begin{equation}
\Revision{P_\Lambda}=\begin{bmatrix}0_{km\times r} & I_{km}\end{bmatrix},
\end{equation}
so $\Revision{\boldsymbol{\ell}}=\Revision{P_\Lambda} \mathbf{q}=\mathrm{vec}(\Revision{\Lambda})$.
Let $\boldsymbol{\alpha}_t=\Revision{\Lambda}\mathbf s_t\in\mathbb{R}^k$, and let $j^\star$ denote the index of the local basis corresponding to the user-selected region. Then $S$ is the selection matrix such that
$S\Revision{\boldsymbol{\ell}}=[\boldsymbol{\alpha}_{1,j^\star},\dots,\boldsymbol{\alpha}_{T,j^\star}]^\top\in\mathbb{R}^T$,
and $\mathbf{a}\in\mathbb{R}^T$ is the target temporal profile for this selected basis (e.g., a wrapped Gaussian around the target frame).
The optimization is
\begin{equation}
\begin{aligned}
\min_{\mathbf{q}}\;&
w_{\mathrm{red}}\|\Revision{A_{\mathrm{loc}}}\mathbf{q}-\mathbf{y}\|_2^2
+w_u\|\Revision{P_\Lambda} \mathbf{q}\|_2^2 \\
&+w_{\mathrm{profile}}\|S\Revision{P_\Lambda} \mathbf{q}-\mathbf{a}\|_2^2 \\
\text{s.t.}\;&
\Revision{A_{\mathrm{cl}}}\mathbf{q}=0.
\end{aligned}
\end{equation}

\paragraph{KKT linear system.}
This constrained quadratic problem has the form
\begin{equation}
\min_{\mathbf{q}}\;\frac12 \mathbf{q}^\top \Revision{Q_{\mathrm{loc}}} \mathbf{q}-\mathbf{b}_q^\top \mathbf{q}
\quad
\text{s.t.}\quad
\Revision{A_{\mathrm{cl}}}\mathbf{q}=0,
\end{equation}
with
\begin{equation}
\Revision{Q_{\mathrm{loc}}}=
2w_{\mathrm{red}}\Revision{A_{\mathrm{loc}}}^\top \Revision{A_{\mathrm{loc}}}
+2w_u\Revision{P_\Lambda}^\top \Revision{P_\Lambda}
+2w_{\mathrm{profile}}\Revision{P_\Lambda}^\top S^\top S \Revision{P_\Lambda},
\end{equation}
\begin{equation}
\mathbf{b}_q=
2w_{\mathrm{red}}\Revision{A_{\mathrm{loc}}}^\top \mathbf{y}
+2w_{\mathrm{profile}}\Revision{P_\Lambda}^\top S^\top \mathbf{a}.
\end{equation}
Therefore,
\begin{equation}
\begin{bmatrix}
\Revision{Q_{\mathrm{loc}}} & \Revision{A_{\mathrm{cl}}}^\top\\
\Revision{A_{\mathrm{cl}}} & 0
\end{bmatrix}
\begin{bmatrix}
\mathbf{q}\\ \boldsymbol{\nu}
\end{bmatrix}
=
\begin{bmatrix}
\mathbf{b}_q\\ 0
\end{bmatrix}.
\end{equation}
After solving, recover $\tilde{\mathbf z}_1$ and $\Revision{\Lambda}$ from $\mathbf{q}$, then roll out
$\tilde{\mathbf z}_{t+1}=\hat{\mathbf K}\tilde{\mathbf z}_t+\Revision{B}\Revision{\Lambda}\mathbf s_t$.


\section{Additional Results}
\label{sec:additional_results}

\begin{figure*}[t]
    \centering
    \includegraphics[width=\textwidth]{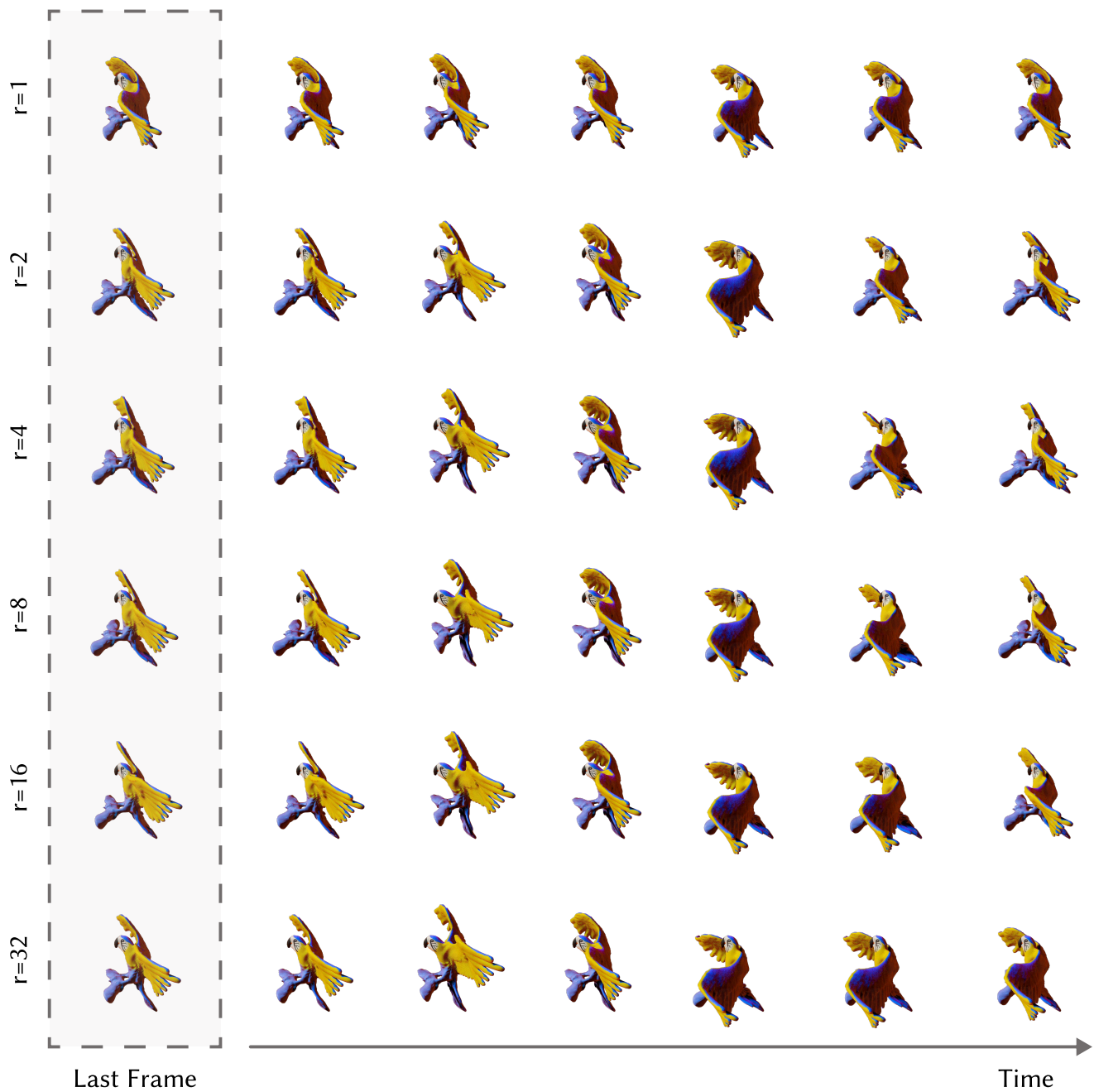}
    \caption{\Revision{Effect of subspace dimension. Cyclic edits of the parrot sequence using reduced spaces of rank $(r=1,2,4,8,16,32)$. The dashed column shows the last frame of each edited sequence, which coincides with the first frame by construction. Increasing the rank recovers more high frequency details}}
    \label{fig:parrot_rank_figures}
\end{figure*}
\begin{figure*}
    \centering
    \includegraphics[width=\textwidth]{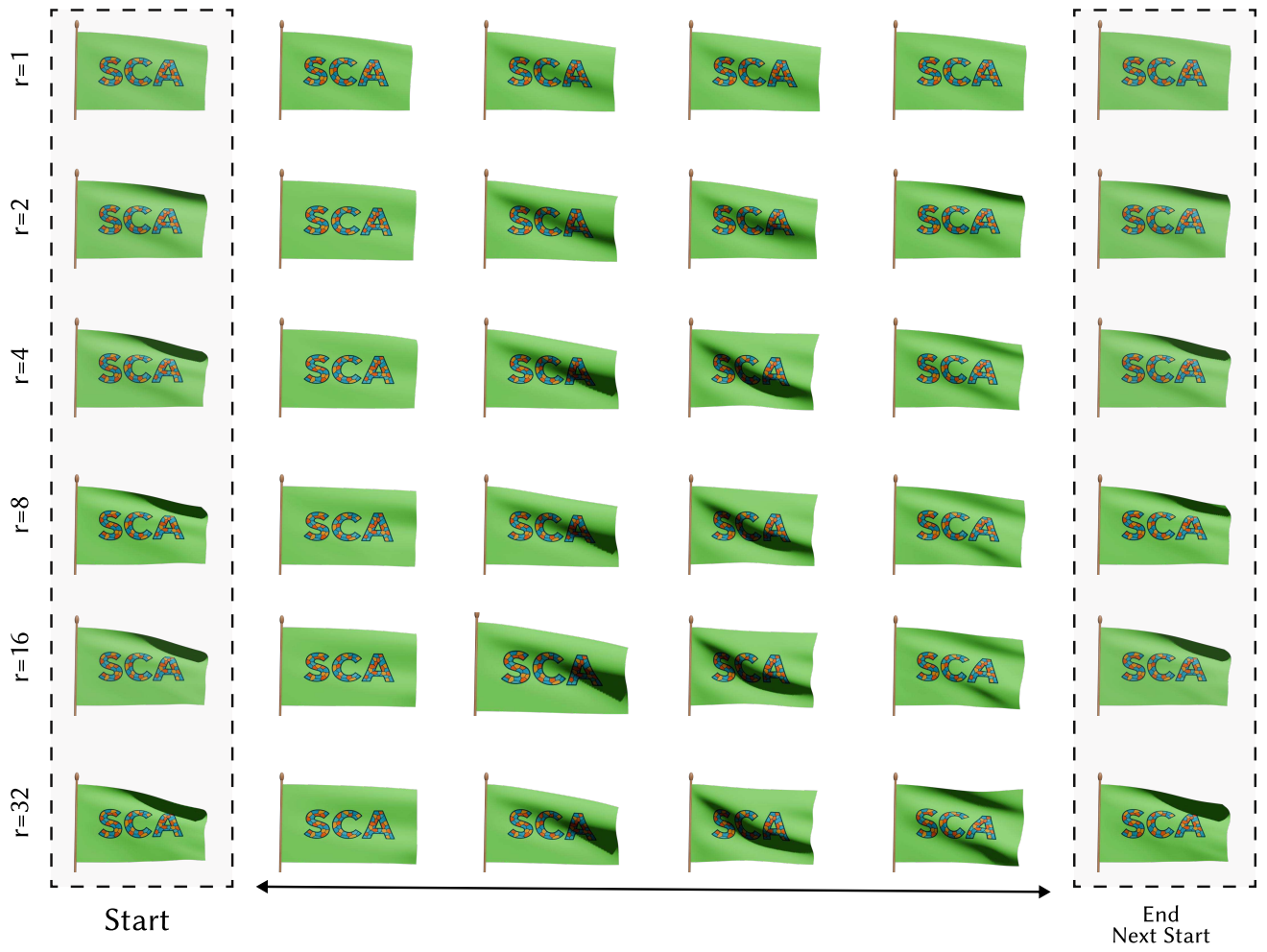}
    \caption{\Revision{Effect of subspace dimension. Cyclic edits of the flag sequence using reduced spaces of rank $(r=1,2,4,8,16,32)$. The dashed column shows the last frame and start frames of each edited sequence, which agree numerically. Increasing the rank recovers more high frequency details}}
    \label{fig:flag_rank_figures}
\end{figure*}

\clearpage

\end{document}